\documentclass[10pt]{article}

\usepackage{amsmath}
\usepackage{amsfonts}
\usepackage[final]{epsfig}
\usepackage{times}

\newcommand{\Ai}{\operatorname{Ai}}
\newcommand{\Bi}{\operatorname{Bi}}
\newcommand{\Ci}{\operatorname{Ci}}
\newcommand{\Qk}{\operatorname{Q}}
\newcommand{\Qi}{\operatorname{Qi}}
\newcommand{\Zi}{\operatorname{Zi}}
\newcommand{\zi}{\operatorname{{\mathfrak Z}i}}
\newcommand{\PP}{\operatorname{PP}}
\newcommand{\Klm}{\operatorname{K}}
\newcommand{\etalchar}[1]{$^{#1}$}

\begin{document}

\begin{center}
{\LARGE Ballistic matter waves with angular momentum:\\
Exact solutions and applications\\}
{\Large\ \\Christian Bracher\footnote{E-mail: {\tt cbracher@fizz.phys.dal.ca}}$^{,1}$,
Tobias Kramer$^2$ and Manfred Kleber$^2$\\}
\end{center}

\begin{center}
$^1$Department of Physics and Atmospheric Science, Dalhousie University,\\
Halifax, N.S.\ B3H~3J5, Canada\\
$^2$Physik--Department T30c, Technische Universit\"at M\"unchen,\\
James--Franck--Str., 85747 Garching, Germany
\end{center}

\begin{center}
\large August 22, 2002
\end{center}

\bigskip

\noindent {\small \bf Abstract.\rm \ \
An alternative description of quantum scattering processes rests on inhomogeneous terms amended to the Schr\"odinger equation.  We detail the structure of sources that give rise to multipole scattering waves of definite angular momentum, and introduce pointlike multipole sources as their limiting case.  Partial wave theory is recovered for freely propagating particles.  We obtain novel results for ballistic scattering in an external uniform force field, where we provide analytical solutions for both the scattering waves and the integrated particle flux.  Our theory directly applies to $p$--wave photodetachment in an electric field.  Furthermore, illustrating the effects of extended sources, we predict some properties of vortex-bearing atom laser beams outcoupled from a rotating Bose--Einstein condensate under the influence of gravity.}

\bigskip

\noindent {\small \bf PACS numbers.\rm \quad
\vtop{%
\hbox{{\tt 03.65.Nk} --- Scattering theory\hss}
\hbox{{\tt 03.75.-b} --- Matter waves\hss}
\hbox{{\tt 03.75.Fi} --- Phase coherent atomic ensembles; quantum condensation phenomena\hss}
\hbox{{\tt 32.80.Gc} --- Photodetachment of atomic negative ions\hss}}}

\bigskip

\section{Introduction}
\label{sec:Intro}

The customary approach to elastic quantum scattering phenomena employs a superposition of an incoming plane wave and an outbound scattering wave that emerges from a localized scattering potential.  It is sometimes advisable to reduce the complexity of this process by dividing the scattering event into subsequent ``absorption'' and ``emission'' stages.  The evolution of the emerging wave is then considered separately.  Obviously, in this description a ``reservoir'' of particles in the interaction region is required that continuously feeds the stationary scattering wave.  Since the particle number is a conserved quantity in the standard quantum picture, we devise a modified approach:  In analogy to electrodynamics and other field theories, an inhomogeneous ``source term'' added to the Schr\"odinger equation allows for particle generation in a finite volume.  It was Schwinger who introduced the idea of particle sources in field theory. In doing so, Schwinger could avoid the use of operator fields. Interestingly enough, he also presented the non-relativistic limit of such a particle source \cite{Schwinger1973a}.  Recent examples illustrating the use of the source formalism are presented in Ref.~\cite{Bracher1997a,Kramer2002a}.

In our contribution, we inquire into quantum sources that give rise to scattering waves carrying non-vanishing angular momentum, which we will denote as multipole waves.  In the long wave (or low energy) limit, apart from its angular dependence the actual structure of the source becomes insignificant, and the scattering process may be properly modelled using an idealized pointlike source of suitable orbital symmetry.  Technically, these ``multipole sources'' are obtained from the Dirac $\delta$--distribution (that itself pertains to isotropic or s--wave emission, see Ref.~\cite{Kramer2002a}) by a simple differentiation procedure outlined in Section~\ref{sec:Multi} that grants immediate access to the corresponding multipole wave and currents.

For scattering waves propagating freely (or in a central potential), multipole sources generate the spherical waves familiar from partial wave theory (Section~\ref{sec:Multi}).  Non-trivial results emerge, however, when the scattered particles are subject to acceleration in a homogeneous force field \cite{Bracher1998a,Kramer2002a}.  We present analytical expressions for the ensuing ballistic multipole waves and currents that so far have not appeared in the literature, and discuss some of their intriguing features in Section~\ref{sec:Ball}.  These developments are directly applicable towards near-threshold photodetachment microscopy, an experimental technique recently introduced by Blondel et~al.\ \cite{Blondel1996a,Blondel1999a} that allows to observe interference of electron waves on a macroscopic scale.  Here, we predict the photoelectron distribution in p--wave detachment (Section~\ref{sec:Photo}).
In general, the multipole formalism breaks down when the spatial extension of the source becomes comparable to the particle wavelength.  In the ballistic environment, Gaussian sources provide an important exception since the scattering waves generated by them apparently converge onto a displaced pointlike ``virtual source'' \cite{Kramer2002a}.  Actually, this situation is encountered for an atom laser beam outcoupled from an ideal Bose--Einstein condensate (BEC) that is subsequently accelerated in the earth's gravitational field \cite{Mewes1997a,Bloch1999a}.  Angular momentum transfer to the superfluid condensate leads to the formation of vortices that in turn act as sources for higher modes of the resulting atom laser.  The effects of vortices on the beam profile are investigated in Section~\ref{sec:Atom}.  Exact solutions are presented for a single vortex in an otherwise spherically symmetric ideal BEC.  Furthermore, we calculate the structures imprinted on the atomic beam by the rotating vortex lattices recently realized experimentally \cite{Madison2000a,Aboshaeer2001a,Engels2002a}.  Finally, some useful mathematical developments have been assembled in the appendix.

\section{Multipole sources}
\label{sec:Multi}

Let us start out with a brief overview of the source formalism and its basic results, emphasizing its kinship to conventional scattering theory.  Throughout much of this section, we will study the behaviour of idealized ``multipole sources'' that emerge in the long-wave limit.  For freely propagating particles, the connection to the partial wave formalism is straightforward.

\subsection{A primer on source theory}
\label{sec:Multi1}

In order to motivate the introduction of quantum sources, we investigate elastic potential scattering as a tutorial example \cite{Messiah1964a}.  (The source approach is readily extended to more sophisticated problems like the atom laser (Section~\ref{sec:Atom}), as illustrated in Ref.~\cite{Kramer2002a}.)  In the customary treatment, the total wave function $\psi(\mathbf r)$ in the potential $V(\mathbf r)$ is decomposed into an incoming wave $\psi_{\rm in}(\mathbf r)$, usually a plane wave, and a scattered wave $\psi_{\rm sc}(\mathbf r)$ that may be written as a sum of partial waves $\psi_{lm}(\mathbf r)$ of definite spherical symmetry:  $\psi(\mathbf r) = \psi_{\rm in}(\mathbf r) + \psi_{\rm sc}(\mathbf r)$. Obviously, $\psi_{\rm in}(\mathbf r)$ is not an eigenfunction to the full Hamiltonian $H = T + V$, but rather to a simpler ``unperturbed'' Hamiltonian $H_0 = T + U$: $H_0\psi_{\rm in}(\mathbf r) = E\psi_{\rm in}(\mathbf r)$.  (Often, one sets $U = 0$.  When long-range forces are present, like in Coulomb scattering \cite{Messiah1964a}, this choice is poor, and $U(\mathbf r)$ should account for the interaction potential.  See also Section~\ref{sec:Ball}.)  Consequently, this procedure leads to the introduction of the scattering potential $W(\mathbf r) = V(\mathbf r) - U(\mathbf r)$, and the stationary Schr\"odinger equation reads, as usual, $[E-H]\psi(\mathbf r) = 0$ or:
\begin{equation}
\label{eq:Multi1.1}
\left[ E - H_0 \right] [\psi_{\rm in}(\mathbf r) + \psi_{\rm sc}(\mathbf r)] = W(\mathbf r) [\psi_{\rm in}(\mathbf r) + \psi_{\rm sc}(\mathbf r)] \;.
\end{equation}
Since $\psi_{\rm in}(\mathbf r)$ is an eigenfunction to $H_0$, we may state (\ref{eq:Multi1.1}) in the equivalent form:
\begin{equation}
\label{eq:Multi1.2}
\left[ E - H_0 - W(\mathbf r) \right] \psi_{\rm sc}(\mathbf r) = W(\mathbf r) \psi_{\rm in}(\mathbf r) \;.
\end{equation}
We infer that the scattering wave $\psi_{\rm sc}(\mathbf r)$ solves a Schr\"odinger equation for the full Hamiltonian $H = H_0 + W(\mathbf r)$, albeit with an additional inhomogeneous term $\sigma(\mathbf r) = W(\mathbf r) \psi_{\rm in}(\mathbf r)$ present.

In view of other inhomogeneous field equations, e.~g.\ Maxwell's equations, the right-hand term $\sigma(\mathbf r)$ in (\ref{eq:Multi1.2}) is identified as a source for the scattering wave $\psi_{\rm sc}(\mathbf r)$.  This observation motivates a simple picture for the scattering process:  The incoming wave $\psi_{\rm in}(\mathbf r)$, via the perturbation $W(\mathbf r)$, feeds particles into the scattering wave $\psi_{\rm sc}(\mathbf r)$ that is governed by the Hamiltonian $H$.  Thus, the decomposition of the wave function into an incoming and a scattered part naturally leads to the notion of a quantum source.

We now turn to the mathematical aspects of (\ref{eq:Multi1.2}).  Introducing the energy Green function $G(\mathbf r,\mathbf r';E)$ for the Hamiltonian $H$ defined via \cite{Economou1983a}:
\begin{equation}
\label{eq:Multi1.3}
\left[ E - H_0 - W(\mathbf r) \right] G(\mathbf r,\mathbf r';E) = \delta(\mathbf  r - \mathbf r') \;,
\end{equation}
a solution to (\ref{eq:Multi1.2}) in terms of a convolution integral reads:
\begin{equation}
\label{eq:Multi1.4}
\psi_{\rm sc}(\mathbf r) = \int {\rm d}^3r'\, G(\mathbf r,\mathbf r';E) \sigma(\mathbf r') \;.
\end{equation}
In general, this result is not unique. However, any two solutions $\psi_{\rm sc}(\mathbf r)$ differ only by an eigenfunction $\psi_{\rm hom}(\mathbf r)$ of $H$:  $H \psi_{\rm hom}(\mathbf r) = E \psi_{\rm hom}(\mathbf r)$.  The ambiguity in $\psi_{\rm sc}(\mathbf r)$ is resolved by the demand that $G(\mathbf r,\mathbf r';E)$ presents a retarded solution characterized by outgoing-wave behaviour as $r\rightarrow\infty$.  Formally, this enforces the choice \cite{Halperin1952a}:
\begin{equation}
\label{eq:Multi1.5}
G(\mathbf r,\mathbf r';E) = \lim_{\eta \rightarrow 0^+} \left\langle \mathbf r \left| \frac1{E-H+{\rm i}\eta} \right| \mathbf r' \right\rangle = \left\langle \mathbf r \left| \PP \left( \frac1{E-H} \right) - {\rm i}\pi\delta(E-H) \right| \mathbf r' \right\rangle \;,
\end{equation}
where $\PP(\ldots)$ denotes the Cauchy principal value of the energy integration.

Defining the current density in the scattering wave in the usual fashion by $\mathbf j(\mathbf r) = \hbar \Im[\psi_{\rm sc}(\mathbf r)^* \boldsymbol\nabla \psi_{\rm sc}(\mathbf r)]/M$ (where for simplicity we omitted the vector potential $\mathbf A(\mathbf r)$, see \cite{Kramer2001a}), the inhomogeneous Schr\"odinger equation (\ref{eq:Multi1.2}) gives rise to a modified equation of continuity \cite{Bracher1997a,Kramer2002a}:
\begin{equation}
\label{eq:Multi1.6}
\boldsymbol\nabla\cdot\mathbf j(\mathbf r) = - \frac2\hbar \Im\left[ \sigma(\mathbf r)^* \psi_{\rm sc}(\mathbf r) \right] \;.
\end{equation}
Thus, the inhomogeneity $\sigma(\mathbf r)$ also acts as a source for the particle current $\mathbf j(\mathbf r)$.  By integration over the source volume, and inserting (\ref{eq:Multi1.4}), we obtain a bilinear expression for the total particle current $J(E)$, i.~e., the total scattering rate:
\begin{equation}
\label{eq:Multi1.7}
J(E) = - \frac2\hbar \Im\left[ \int {\rm d}^3r \int {\rm d}^3r' \sigma(\mathbf r)^* G(\mathbf r,\mathbf r';E) \sigma(\mathbf r') \right] \;.
\end{equation}

Some important identities concerning the total current $J(E)$ are most easily recognized in a formal Dirac bra-ket representation.  In view of (\ref{eq:Multi1.5}), we may express $J(E)$ by:
\begin{equation}
\label{eq:Multi1.8}
J(E) = -\frac2\hbar \Im\left[ \left\langle \sigma \left| G \right| \sigma \right\rangle \right] = \frac{2\pi}\hbar \left\langle \sigma \left| \delta(E-H) \right| \sigma \right\rangle \;,
\end{equation}
from which the sum rule immediately follows \cite{Kramer2002a}:
\begin{equation}
\label{eq:Multi1.9}
\int_{-\infty}^{\infty} {\rm d}E\, J(E) = \frac{2\pi}\hbar \langle\sigma|\sigma\rangle = \frac{2\pi}\hbar \int {\rm d}^3r |\sigma(\mathbf r)|^2 \;,
\end{equation}
(provided this integral exists).  In order to connect (\ref{eq:Multi1.7}) to the findings of conventional scattering theory, we display $J(E)$ in an entirely different, yet wholly equivalent fashion.  Employing a complete orthonormal set of eigenfunctions $| \psi_{\rm fi} \rangle$ of the Hamiltonian $H$, $\delta(E-H)| \psi_{\rm fi} \rangle = \delta(E-E_{\rm fi})| \psi_{\rm fi} \rangle$ follows, and replacing $| \sigma \rangle = W| \psi_{\rm in} \rangle$ (\ref{eq:Multi1.5}), we may formally decompose (\ref{eq:Multi1.8}) into a sum over eigenfunctions:
\begin{equation}
\label{eq:Multi1.10}
J(E) = \frac{2\pi}\hbar \sum_{\rm fi} \delta(E-E_{\rm fi}) \left| \left\langle \psi_{\rm fi} |W| \psi_{\rm in} \right\rangle \right|^2 \;.
\end{equation}
Thus, Fermi's golden rule is recovered.  Another noteworthy consequence of (\ref{eq:Multi1.7}) and (\ref{eq:Multi1.8}) emerges in the limit of pointlike sources, $\sigma(\mathbf r) \sim C \delta(\mathbf r-\mathbf R)$.  We then find \cite{Bracher1997a}:
\begin{equation}
\label{eq:Multi1.11}
J(E) = -\frac2\hbar |C|^2 \Im[G(\mathbf R,\mathbf R;E)] = \frac{2\pi}\hbar |C|^2 n(\mathbf R;E) \;,
\end{equation}
where $n(\mathbf R;E) = \sum_{\rm fi} \delta(E-E_{\rm fi}) |\psi_{\rm fi}(\mathbf R)|^2$ is the local density of states of $H$ at the source position $\mathbf R$.  Equation (\ref{eq:Multi1.11}) forms the theoretical basis of the Tersoff--Hamann description of scanning tunneling microscopy \cite{Tersoff1983a,Bracher1997a}.

\subsection{Multipole sources, waves, and currents}
\label{sec:Multi2}

Evidently, the theory of quantum sources becomes particularly simple for point-like sources.  In this case, the otherwise bothersome integrations involved in the determination of the scattering wave $\psi_{\rm sc}(\mathbf r)$ (\ref{eq:Multi1.4}) and the total current $J(E)$ (\ref{eq:Multi1.7}) become trivial.  For the naive choice of a point source, $\sigma (\mathbf r) = C \delta(\mathbf r-\mathbf R)$, $\psi_{\rm sc}(\mathbf r)$ is simply proportional to the Green function $G(\mathbf r,\mathbf R; E)$ itself, and $J(E)$ follows from the Tersoff--Hamann rule (\ref{eq:Multi1.11}).  The approximation of a pointlike source is obviously well justified in near-threshold scattering $(E\rightarrow 0)$, where the long wavelength of the emerging wave effectively obliterates the internal structure of the source.  This statement, however, must be taken \textsl{cum grano salis}, for it does not take into account the orbital structure of the scattering wave.

In fact, as discussed below, a simple delta source $\sigma(\mathbf r) \sim \delta(\mathbf r-\mathbf r')$ invariably leads to a locally isotropic emission pattern, i.~e., describes scattering into an $s$--wave.  Despite often being appropriate in practice \cite{Blondel1996a,Blondel1999a,Kramer2001a}, conservation of angular momentum may enforce selection rules that restrict scattering to higher multipole waves.  For these, the idealized point source approach must be suitably modified.  We proceed by analogy with the well-established multipole formalism in potential theory (or electrostatics) as another inhomogeneous field equation.

Like the scattering wave for a simple point source, the Green function of potential theory equals the field created by a source of unit strength located at $\mathbf r'$, $G(\mathbf r,\mathbf r') = - 1/4\pi|\mathbf r-\mathbf r'|$.  Additional solutions that likewise show a singularity at $\mathbf r'$ can be constructed via differentiation with respect to the source position $\mathbf r' = (x', y', z')$.  Of special significance are the multipole potentials $\Phi_{lm}(\mathbf r,\mathbf r')$:
\begin{equation}
\label{eq:Multi2.1}
\Phi_{lm}(\mathbf r,\mathbf r') = \frac{Y_{lm}(\hat e_{\mathbf r - \mathbf r'})}{|\mathbf r-\mathbf r'|^{l+1}}
= \frac{\Klm_{lm}(\mathbf r - \mathbf r')}{|\mathbf r-\mathbf r'|^{2l+1}} \;,
\end{equation}
since they are obviously of $(l,m)$ spherical symmetry.  Here, we introduced the harmonic polynomials $\Klm_{lm}(\mathbf r) = r^l Y_{lm}(\hat r)$, homogeneous polynomials of order $l$ in the coordinates $x,y,z$ that are eigenfunctions of the angular momentum operator \cite{Hobson1931a, MorseFeshbach1953a,Mueller1966a}.  Interestingly, the same polynomial in momentum space, known as the spherical tensor gradient $\Klm_{lm}(\boldsymbol\nabla') = \Klm_{lm}(\partial_{x'}, \partial_{y'}, \partial_{z'})$ \cite{Bayman1978a,Rowe1978a,Weniger1983a}, can be employed to generate the multipole potentials from the Green function $G(\mathbf r,\mathbf r')$:
\begin{equation}
\label{eq:Multi2.2}
\Klm_{lm}(\boldsymbol\nabla') G(\mathbf r,\mathbf r') = - \frac{(2l-1)!!}{4\pi} \Phi_{lm}(\mathbf r,\mathbf r') \;.
\end{equation}
Thus, the spherical tensor gradient imprints the orbital structure onto $G(\mathbf r,\mathbf r')$.  Since $\Delta G(\mathbf r,\mathbf r') = \delta(\mathbf r-\mathbf r')$ holds, we formally obtain from (\ref{eq:Multi2.2}):
\begin{equation}
\label{eq:Multi2.3}
\Delta\Phi_{lm}(\mathbf r,\mathbf r') = - \frac{4\pi}{(2l-1)!!} \Klm_{lm}(\boldsymbol\nabla') \delta(\mathbf r-\mathbf r') \;.
\end{equation}
(Note that the differentiation proceeds with respect to $\mathbf r'$.)  Apart from prefactors, the multipole potentials $\Phi_{lm}(\mathbf r,\mathbf r')$ are thus generated by the spherical tensor gradients of the delta distribution.

Accordingly, for the purpose of the quantum source problem, we define multipole point sources $\delta_{lm}(\mathbf r-\mathbf r')$ via \cite{Bayman1978a,Rowe1978a}:
\begin{equation}
\label{eq:Multi2.4}
\delta_{lm}(\mathbf r-\mathbf r') = \Klm_{lm}(\boldsymbol\nabla') \delta(\mathbf r-\mathbf r') \;.
\end{equation}
Since $\Klm_{lm}(\boldsymbol\nabla')$ and the Hamilton operator $H(\mathbf r,\mathbf p)$ always commute, the inhomogeneous Schr\"odinger equation:
\begin{equation}
\label{eq:Multi2.5}
[E - H] G_{lm}(\mathbf r,\mathbf r'; E) = \delta_{lm}(\mathbf r-\mathbf r')
\end{equation}
is formally solved by the multipole Green functions $G_{lm}(\mathbf r,\mathbf r'; E)$ that are available from $G(\mathbf r,\mathbf r'; E)$ by differentiation:
\begin{equation}
\label{eq:Multi2.6}
G_{lm}(\mathbf r,\mathbf r'; E) = \Klm_{lm}(\boldsymbol\nabla') G(\mathbf r,\mathbf r'; E) \;.
\end{equation}
(We remark in passing that $G(\mathbf r,\mathbf r'; E)$ differs from the $s$--wave multipole Green function $G_{00}(\mathbf r,\mathbf r'; E)$ only by a factor $\sqrt{4\pi}$.)  The multipole point sources $\delta_{lm}(\mathbf r-\mathbf r')$ (\ref{eq:Multi2.4}) and Green functions (\ref{eq:Multi2.6}) provide the extension of the convenient multipole formalism towards sources with internal orbital structure.  For freely propagating particles $[U(\mathbf r) = 0]$, they naturally emerge in the partial wave decomposition of the scattering wave.  (See Section~\ref{sec:Multi4}.)

Next, we turn our attention to the currents generated by multipole point sources.  Assuming a superposition of several of these sources at a fixed location $\mathbf r'$, $\sigma(\mathbf r) = \sum_{lm} \lambda_{lm} \delta_{lm}(\mathbf r-\mathbf r')$, the resulting scattering wave will read:
\begin{equation}
\label{eq:Multi2.7}
\psi_{\rm sc}(\mathbf r) = \sum_{lm} \lambda_{lm} G_{lm}(\mathbf r,\mathbf r';E) \;.
\end{equation}
The current density $\mathbf j(\mathbf r)$ due to this wave function then may be expressed as a bilinear form in the amplitudes $\lambda_{lm}$:
\begin{equation}
\label{eq:Multi2.8}
\mathbf j(\mathbf r) = \sum_{lm} \sum_{l'm'} \lambda_{lm}^*\, {\mathbf j}_{lm,l'm'}(\mathbf r) \lambda_{l'm'} \;,
\end{equation}
where the vectors ${\mathbf j}_{lm,l'm'}(\mathbf r)$, given by:
\begin{equation}
\label{eq:Multi2.9}
{\mathbf j}_{lm,l'm'}(\mathbf r) = -\frac{{\rm i}\hbar}{2M}
\left\{ G_{lm}^*(\mathbf r,\mathbf r'; E) \boldsymbol\nabla G_{l'm'}(\mathbf r,\mathbf r'; E) -
G_{l'm'}(\mathbf r,\mathbf r'; E) \boldsymbol\nabla G_{lm}^*(\mathbf r,\mathbf r'; E) \right\} \;,
\end{equation}
form the elements of the hermitian current density matrix.

Of particular interest is the total current $J(E)$ carried by the scattering wave (\ref{eq:Multi2.7}), i.~e., the integrated current density $\mathbf j(\mathbf r)$ (\ref{eq:Multi2.8}).  Inserting the corresponding multipole source into the general current formula (\ref{eq:Multi1.7}), after integration by parts we first obtain:
\begin{equation}
\label{eq:Multi2.10}
J(E) = - \frac2\hbar \lim_{\mathbf r\rightarrow \mathbf r'} \Im\left[
\sum_{lm} \sum_{l'm'} \lambda_{lm}^* \lambda_{l'm'}
\Klm_{lm}^*(\boldsymbol\nabla) \Klm_{lm}(\boldsymbol\nabla') G(\mathbf r,\mathbf r'; E) \right] \;.
\end{equation}
Thus, total multipole currents can be extracted from the Green function $G(\mathbf r,\mathbf r'; E)$ through differentiation with respect to $\mathbf r$ as well as $\mathbf r'$ and a subsequent limiting procedure, i.~e., a number of elementary operations.  We may further transform (\ref{eq:Multi2.10}) to obtain a more symmetrical representation of the current that again involves a hermitian matrix $J_{lm,l'm'}(E)$ [cf.~(\ref{eq:Multi2.8})]:
\begin{equation}
\label{eq:Multi2.11}
J(E) = \sum_{lm} \sum_{l'm'} \lambda_{lm}^* J_{lm,l'm'}(E) \lambda_{l'm'} \;,
\end{equation}
where the components of the total multipole current matrix $J_{lm,l'm'}(E)$ are given by:
\begin{equation}
\label{eq:Multi2.12}
J_{lm,l'm'}(E) = \frac{\rm i}\hbar \lim_{\mathbf r\rightarrow \mathbf r'}
\Klm_{lm}^*(\boldsymbol\nabla) \Klm_{lm}(\boldsymbol\nabla')
\left\{ G(\mathbf r,\mathbf r'; E) - G(\mathbf r',\mathbf r; E)^* \right\} \;.
\end{equation}
For simplicity, we will denote the (real and positive) diagonal elements of this matrix as the $(l,m)$ multipole currents $J_{lm}(E)$:  $J_{lm}(E) = J_{lm,lm}(E)$.

If the Hamiltonian $H$ possesses elements of angular symmetry, the number of non-vanishing total current matrix elements $J_{lm,l'm'}(E)$ (\ref{eq:Multi2.12}) is greatly reduced.  Assume that the generator of a rotation $L$ commutes with $H$;  then, it will also commute with the resolvent operator $G = [E - H + {\rm i}\eta]^{-1}$.  Consequently, if the source states $|\sigma\rangle$ and $|\sigma'\rangle$ are eigenstates of $L$ with different eigenvalues, the mixed matrix element $\langle \sigma |G| \sigma' \rangle$, and hence its contribution to the total current $J(E)$ in (\ref{eq:Multi1.8}), is bound to vanish.  This implies that in case of full rotational symmetry of $H$, e.~g.\ for a freely propagating wave with $U(\mathbf r) = 0$, the total multipole current matrix (\ref{eq:Multi2.12}) is diagonal, $J_{lm,l'm'}(E) = J_{lm}(E) \delta_{ll'} \delta_{mm'}$, and the total current in (\ref{eq:Multi2.11}) becomes a simple sum:  $J(E) = \sum_{lm} |\lambda_{lm}|^2 J_{lm}(E)$ (Section~\ref{sec:Multi4}).  If the potential $U(\mathbf r)$ is invariant merely with respect to rotations around the $z$--axis (like in the ballistic problem discussed in detail in Section~\ref{sec:Ball}--\ref{sec:Atom}), orthogonality with respect to different values of $m$ prevails:  $J_{lm,l'm'}(E) = 0$ for $m \neq m'$.

Finally, we set out to extend the Tersoff-Hamann description (\ref{eq:Multi1.11}) of the current to cover the case of multipole point sources.  According to (\ref{eq:Multi1.7}), the current $J_{lm}(E)$ for a source located at $\mathbf R$ is formally given by:  $J_{lm}(E) = 2\pi \langle \mathbf R | \Klm_{lm}^*(\mathbf p) \delta(E - H) \Klm_{lm}(\mathbf p) | \mathbf R \rangle / \hbar$.  Expanding this expression again into a complete orthonormal set of eigenfunctions $|\psi_{\rm fi}\rangle$ of $H$, we find:
\begin{equation}
\label{eq:Multi2.13}
J_{lm}(E) = \frac{2\pi}\hbar \sum_{\rm fi} \delta(E-E_{\rm fi}) |\Klm_{lm}(\boldsymbol\nabla) \psi_{\rm fi}(\mathbf R)|^2 \;.
\end{equation}
Therefore, the multipole currents $J_{lm}(E)$ (\ref{eq:Multi2.12}) are proportional to the local density of the respective spherical tensor gradients of the eigenstates of the Hamiltonian $H$ at energy $E$ and position $\mathbf R$.

\subsection{Analytical properties of multipole Green functions}
\label{sec:Multi3}

Our definition of multipole point sources $\delta_{lm}(\mathbf r-\mathbf r')$ (\ref{eq:Multi2.4}) and corresponding Green functions $G_{lm}(\mathbf r,\mathbf r';E)$ (\ref{eq:Multi2.6}) by means of a differentiation procedure is a purely formal one.  Thus, it appears appropriate to corroborate their interpretation as sources of particles with definite angular momentum.  Since we did not assume that the external potential $V(\mathbf r)$, and therefore the Hamiltonian $H$, commute with the angular momentum operators, the rotational symmetry of the multipole Green functions will generally be broken.  However, in the immediate neighbourhood of the source, $\mathbf r\rightarrow \mathbf r'$, the typical angular variation should manifest itself.

Hence, in this section we study the local properties of $G_{lm}(\mathbf r,\mathbf r';E)$ in the vicinity of the point source and identify a universal behaviour that holds under quite general circumstances, viz., under the weak assumption that the potential $V(\mathbf r)$ is a locally analytic function.  Unlike their global structure, the local characteristics of wave functions apparently have found little attention in the physics literature.  The regularity properties of solutions to partial differential equations have been examined thoroughly, however, and the interested reader is referred to the monograph by H{\"o}rmander \cite{Hoermander1963a}.  The subsequent results follow from an adaptation of this formalism to the multipole Schr{\"o}dinger equation (\ref{eq:Multi2.5}) recently performed by one of the authors \cite{Bracher1999a}.

Mathematically spoken, the Hamiltonian operator $H = T + V$ belongs to the class of elliptic operators \cite{Hoermander1963a}, which implies that all solutions of the inhomogeneous problem $[E-H]\psi(\mathbf r) = \sigma(\mathbf r)$ (\ref{eq:Multi1.2}) are analytic in those sectors of space where the potential $V(\mathbf r)$ and the source term $\sigma(\mathbf r)$ are likewise analytic.  It immediately follows that the difference of any two solutions to the point source problems in (\ref{eq:Multi1.3}) and (\ref{eq:Multi2.5}) will be a locally analytic function at $\mathbf r=\mathbf r'$ (provided $V(\mathbf r)$ is regular at $\mathbf r'$).  Yet this statement does not apply to the Green functions themselves, since the source term is singular.  To work around this problem, we employ the results of potential theory (Section~\ref{sec:Multi2}) and eliminate the delta singularity in (\ref{eq:Multi1.3}) by a factorization approach:  We set $\tilde G(\mathbf r,\mathbf r';E) = 2M G(\mathbf r,\mathbf r') f(\mathbf r,\mathbf r';E) / \hbar^2$, where $G(\mathbf r,\mathbf r') = -1/4\pi|\mathbf r-\mathbf r'|$ is the Green function of potential theory.  Then, $f(\mathbf r,\mathbf r';E)$ is a solution to:
\begin{equation}
\label{eq:Multi3.1}
\left[ R^2 \Delta_{\mathbf R} - 2 \mathbf R\cdot\boldsymbol\nabla_{\mathbf R} + R^2 k(\mathbf R)^2 \right]
f(\mathbf r,\mathbf r';E) = 0 \;,
\end{equation}
where $\mathbf R=\mathbf r-\mathbf r'$ and $k(\mathbf R)^2 = 2M[E - V(\mathbf r)]/\hbar^2$.  Additionally, $f(\mathbf r',\mathbf r';E) = 1$ must hold.  In (\ref{eq:Multi3.1}), the source location $\mathbf R = \mathbf o$ is still a special (weakly singular) point, but it can be proven that a unique solution exists that is locally analytic at $\mathbf r=\mathbf r'$ \cite{Bracher1999a}.  Furthermore, this function is real and symmetric $[f(\mathbf r,\mathbf r';E) = f(\mathbf r',\mathbf r;E)]$, and may be represented in the form $(k = k(\mathbf o))$:
\begin{equation}
\label{eq:Multi3.2}
f(\mathbf r,\mathbf r';E) = \cos\left( k|\mathbf r - \mathbf r'| \right) + (\mathbf r-\mathbf r')^2 {\cal O}(\mathbf r,\mathbf r') \;.
\end{equation}
(For free particles, the correction is absent.  See Section~\ref{sec:Multi4}.)  Now, $\tilde G(\mathbf r,\mathbf r';E)$ is not the Green function itself, but differs from it only by an analytic expression $g(\mathbf r,\mathbf r';E)$.  Hence, $G(\mathbf r,\mathbf r';E)$ always may be displayed as a sum involving two regular functions:
\begin{equation}
\label{eq:Multi3.3}
G(\mathbf r,\mathbf r';E) = - \frac M{2\pi\hbar^2} \frac{f(\mathbf r,\mathbf r';E)}{|\mathbf r-\mathbf r'|} +
g(\mathbf r,\mathbf r';E) \;,
\end{equation}
provided $V(\mathbf r)$ is analytic at $\mathbf r=\mathbf r'$.  (Notably, (\ref{eq:Multi3.3}) does not hold for the Green function of the Coulomb problem \cite{Hostler1963a,Hostler1964a}, where a logarithmic singularity is encountered.)  Clearly, as $\mathbf r\rightarrow \mathbf r'$, the Green function (\ref{eq:Multi3.3}) adopts a universal shape regardless of $V(\mathbf r)$ and mimics $G(\mathbf r,\mathbf r')$:  Locally, isotropy prevails.  Obviously, the imaginary part in (\ref{eq:Multi3.3}) remains well-behaved as $\mathbf r\rightarrow \mathbf r'$, thus justifying the expression (\ref{eq:Multi1.11}) for the current $J(E)$ generated by the point source.

It is now but a small step to generalize these results towards the multipole point sources $\delta_{lm}(\mathbf r-\mathbf r')$ (\ref{eq:Multi2.4}).  Applying the spherical tensor gradient to the Green function (\ref{eq:Multi3.3}) according to the rule (\ref{eq:Multi2.6}), we obtain \cite{Bracher1999a}:
\begin{equation}
\label{eq:Multi3.4}
G_{lm}(\mathbf r,\mathbf r';E) = - \frac M{2\pi\hbar^2} \frac{f_{lm}(\mathbf r,\mathbf r';E)}{|\mathbf r-\mathbf r'|^{2l+1}} + g_{lm}(\mathbf r,\mathbf r';E) \;,
\end{equation}
where $g_{lm}(\mathbf r,\mathbf r';E) = \Klm_{lm}(\boldsymbol\nabla') g(\mathbf r,\mathbf r';E)$ and $f_{lm}(\mathbf r,\mathbf r';E)$ are again analytic functions at $\mathbf r=\mathbf r'$, and the series expansion of the latter reads:
\begin{equation}
\label{eq:Multi3.5}
f_{lm}(\mathbf r,\mathbf r';E) = \Klm_{lm}(\mathbf R) \left[ (kR)^{l+1} n_l(kR) \right] + {\cal O}(R^{l+3}) \;.
\end{equation}
Here, $n_l(z)$ denotes the irregular spherical Bessel function of order $l$ \cite{Abramowitz1965a}.  In the limit $\mathbf r\rightarrow \mathbf r'$, the bracketed expression in (\ref{eq:Multi3.5}) tends towards the constant value $(2l-1)!!$, leading to the principal asymptotic form of $G_{lm}(\mathbf r,\mathbf r';E)$ in the vicinity of the source:
\begin{equation}
\label{eq:Multi3.6}
G_{lm}(\mathbf r,\mathbf r';E) \sim - \frac M{2\pi\hbar^2} (2l-1)!! \frac{Y_{lm}(\hat e_{\mathbf r - \mathbf r'})}{|\mathbf r-\mathbf r'|^{l+1}} \;.
\end{equation}
This expression, a multiple of the multipole potentials $\Phi_{lm}(\mathbf r,\mathbf r')$ (\ref{eq:Multi2.1}) of potential theory, is again universal in scope.  (Indeed, the relations (\ref{eq:Multi2.2}) and (\ref{eq:Multi2.3}) are special cases of the general result (\ref{eq:Multi3.4}).)  Obviously, the multipole Green functions (\ref{eq:Multi2.6}) are endowed with the desired property of local orbital symmetry.

\subsection{Example: Free multipole sources}
\label{sec:Multi4}

As an instructive example, we show how the mathematical developments of the previous sections operate for the simple case of freely propagating waves ($U(\mathbf r) = 0$).  Here, the Green function of the scattering problem is just an outgoing spherical wave \cite{Economou1983a}:
\begin{equation}
\label{eq:Multi4.1}
G^{(\rm free)}(\mathbf r,\mathbf r';E) = - \frac M{2\pi\hbar^2}
\frac{\exp({\rm i}k |\mathbf r-\mathbf r'|)}{|\mathbf r-\mathbf r'|} \;,
\end{equation}
with $E = \hbar^2k^2/2M$.  We note here that $G^{(\rm free)}(\mathbf r,\mathbf r';E)$ neatly fits into the general scheme (\ref{eq:Multi3.2}) and (\ref{eq:Multi3.3}), with analytical functions $f(\mathbf r,\mathbf r';E) = \cos kR$, $g(\mathbf r,\mathbf r';E) = -M \sin (kR)/2\pi\hbar^2R$ (with $\mathbf R = \mathbf r-\mathbf r'$).  For our purposes, a well-known expansion of (\ref{eq:Multi4.1}) into spherical Bessel functions is useful \cite{Messiah1964a}:
\begin{equation}
\label{eq:Multi4.2}
G^{(\rm free)}(\mathbf r,\mathbf r';E) = - \frac{2Mk}{\hbar^2}
\sum_{\lambda\mu} j_\lambda(kr_<) h_\lambda^{(+)}(kr_>) Y^*_{\lambda\mu}(\hat r_<) Y_{\lambda\mu}(\hat r_>) \;,
\end{equation}
where $r_< = \min(r,r')$, $r_> = \max(r,r')$, and $j_\lambda(u)$, $h_\lambda^{(+)}(u) = n_\lambda(u) + {\rm i}j_\lambda(u)$ denote the spherical Bessel and Hankel functions of order $\lambda$, respectively.

With the help of (\ref{eq:Multi4.2}), calculation of the multipole Green functions $G_{lm}^{(\rm free)}(\mathbf r,\mathbf r';E)$ (\ref{eq:Multi2.6}) and currents $J_{lm}^{(\rm free)}(E)$ (\ref{eq:Multi2.12}) becomes an easy task.  We exploit translational symmetry and set $\mathbf r'=\mathbf o$.  Then, we may replace $j_\lambda(kr') \sim (kr')^l/(2l+1)!!$ \cite{Abramowitz1965a}, and only the term $(l,m)$ in the sum (\ref{eq:Multi4.2}) survives:
\begin{eqnarray}
\nonumber
G^{(\rm free)}_{lm}(\mathbf R,\mathbf o;E) &=& - \frac{2Mk}{\hbar^2} h_l^{(+)}(kR) Y_{lm}(\hat R)
\left[ \frac{k^l}{(2l+1)!!} \Klm_{lm}(\boldsymbol\nabla') \Klm_{lm}^*(\mathbf r') \right] \\
&=& - \frac {M k^{l+1}}{2\pi\hbar^2} h_l^{(+)}(kR) Y_{lm}(\hat R) \;.
\label{eq:Multi4.3}
\end{eqnarray}
(The term in brackets is independent of $m$ and yields $k^l/4\pi$.)  Again, this expression conforms to the general structure of multipole Green functions laid out in (\ref{eq:Multi3.4}) and (\ref{eq:Multi3.5}).  From (\ref{eq:Multi4.3}), we infer that $G^{(\rm free)}_{lm}(\mathbf r,\mathbf r';E)$ has the spatial symmetry of a spherical wave with $(l,m)$ orbital structure \cite{Messiah1964a}.  Hence, the multipole source approach reproduces here the partial wave formalism of conventional scattering theory.

In a similar vein, we may extract the total multipole current $J_{lm}^{(\rm free)}(E)$ (\ref{eq:Multi2.12}) carried by a freely propagating wave.  The action of the spherical tensor gradient on (\ref{eq:Multi4.3}) yields in the limit $\mathbf r\rightarrow \mathbf o$:
\begin{equation}
\label{eq:Multi4.4}
J_{lm}^{(\rm free)}(E) = \frac{4Mk}{\hbar^3} \left| \frac{k^l}{(2l+1)!!} \Klm_{lm}(\boldsymbol\nabla') \Klm_{lm}^*(\mathbf r') \right|^2 = \frac {M k^{2l+1}}{4\pi^2\hbar^3} \;.
\end{equation}
The characteristic power law dependence of the near-threshold scattering rate on $l$ is known as Wigner's law \cite{Wigner1948a}.

It remains to identify the connection between realistic, extended sources $\sigma(\mathbf r)$ and the idealized point sources $\delta_{lm}(\mathbf r-\mathbf r')$ (\ref{eq:Multi2.4}) introduced in Section~\ref{sec:Multi2}, particularly for near-threshold scattering processes.  Clearly, the replacement of $\sigma(\mathbf r)$ by multipole sources will only be feasible outside the source region, i.~e., in the far-field limit, similar to the use of multipole potentials $\Phi_{lm}(\mathbf r,\mathbf r')$ (\ref{eq:Multi2.1}) in potential theory.  Incidentally, any free-particle source $\sigma(\mathbf r)$ can be substituted by a set of multipole sources $\delta_{lm}(\mathbf r-\mathbf r')$ for an arbitrary choice of the source location $\mathbf r'$, as we are going to show now.  With $\mathbf R=\mathbf r-\mathbf r'$, $\sigma(\mathbf r)$ may be expanded into a spherical series at $\mathbf r'$:
\begin{equation}
\label{eq:Multi4.5}
\sigma(\mathbf r) = \sum_{lm} \Klm_{lm}(\mathbf R) \sigma_{lm}(R) \;,
\end{equation}
with even coefficient functions $\sigma_{lm}(R)$ accessible via projection over spheres centered in $\mathbf r'$.  For further reference, we also define related coefficents $\gamma_{lm}(\mathbf r')$ by:
\begin{equation}
\label{eq:Multi4.6}
\gamma_{lm}(\mathbf r') = \int {\rm d}^3r\, \Klm_{lm}^*(\mathbf r-\mathbf r') \sigma(\mathbf r) =
\int_0^\infty {\rm d}R\, R^{2l+2} \sigma_{lm}(R) \;.
\end{equation}
Inserting the expressions (\ref{eq:Multi4.2}) and (\ref{eq:Multi4.5}) into the convolution integral (\ref{eq:Multi1.4}), and exploiting the orthonormality of the spherical harmonics, we obtain the multipole series for the scattering wave $\psi^{(\rm free)}_{\rm sc}(\mathbf r)$:
\begin{equation}
\label{eq:Multi4.7}
\psi^{(\rm free)}_{\rm sc}(\mathbf r) = - \frac{2Mk}{\hbar^2} \sum_{lm}
\left[ \int_0^\infty {\rm d}R'\, R^{\prime l+2} j_l(kR') \sigma_{lm}(R') \right] h_l^{(+)}(kR) Y_{lm}(\hat R) \;,
\end{equation}
(provided the source is entirely contained in the ball of radius $R'<R$).  By comparison with the free multipole Green function (\ref{eq:Multi4.3}), we find that $\psi^{(\rm free)}_{\rm sc}(\mathbf r) = \sum_{lm} \lambda_{lm}(\mathbf r';E) G^{(\rm free)}_{lm}(\mathbf r,\mathbf r';E)$, where:
\begin{equation}
\label{eq:Multi4.8}
\lambda_{lm}(\mathbf r';E) = \frac{4\pi}{k^l} \int_0^\infty {\rm d}R'\, R^{\prime l+2} j_l(kR') \sigma_{lm}(R') \;.
\end{equation}
Therefore, in the far field limit, the extended source $\sigma (\mathbf r)$ may be replaced by a superposition of multipole sources $\delta_{lm}(\mathbf r-\mathbf r')$ (\ref{eq:Multi2.7}) with strengths $\lambda_{lm}(\mathbf r';E)$ (\ref{eq:Multi4.8}).  In the threshold regime ($E \sim 0$), we may replace the Bessel function in (\ref{eq:Multi4.8}) by its limiting value, which causes the energy dependence to drop from these coefficients:
\begin{equation}
\label{eq:Multi4.9}
\lambda_{lm}(\mathbf r';E = 0) = \frac{4\pi}{(2l+1)!!} \gamma_{lm}(\mathbf r') \;.
\end{equation}
The source components then simply reflect the orbital structure of $\sigma(\mathbf r)$ (\ref{eq:Multi4.6}).  (Although the components $\gamma_{lm}(\mathbf r')$ generally depend on the choice of $\mathbf r'$, the leading non-vanishing coefficients of lowest order $l$ are invariant with respect to shifts of the source location.)

Finally, we take a look at the total current $J^{(\rm free)}(E)$ generated by the source $\sigma(\mathbf r)$.  Starting from the general expression (\ref{eq:Multi1.7}), the expansions (\ref{eq:Multi4.2}) and (\ref{eq:Multi4.5}) yield the simple result:
\begin{equation}
\label{eq:Multi4.10}
J^{(\rm free)}(E) = \frac M{4\pi^2\hbar^3} \sum_{lm} |\lambda_{lm}(\mathbf r';E) |^2  k^{2l+1} =
\sum_{lm} |\lambda_{lm}(\mathbf r';E) |^2 J^{(\rm free)}_{lm}(E) \;,
\end{equation}
[cf.~(\ref{eq:Multi4.4})].  In accordance with the discussion in Section~\ref{sec:Multi2}, the total multipole current matrix $J_{lm,l'm'}^{(\rm free)}(E)$ (\ref{eq:Multi2.11}), (\ref{eq:Multi2.12}) for the free particle problem is diagonal, and $J^{(\rm free)}(E)$ (\ref{eq:Multi4.10}) is simply the properly weighted sum of the multipole currents (\ref{eq:Multi4.4}).  Concluding, we remark that near threshold, higher multipole contributions to $J^{(\rm free)}(E)$ are strongly suppressed.

\section{Ballistic multipole waves and currents}
\label{sec:Ball}

In the preceding section, we applied the multipole wave approach to freely propagating particles.  Obviously, the introduction of this new formalism would be of little merit if it only served to reproduce the well-known results of partial wave decomposition \cite{Messiah1964a}.  Therefore, we now tackle a less trivial problem, the propagation of a multipole scattering wave accelerated by a homogeneous force field $\mathbf F = F\hat e_z$, i.~e., in the presence of a linear potential $U(\mathbf r) = -Fz$.  (Subsequently, we will refer to the dynamics in this environment as quantum ballistic motion.)  As an analytical expression for the corresponding ballistic Green function $G(\mathbf r,\mathbf r';E)$ (\ref{eq:Multi1.3}) is available, the developments of Section~\ref{sec:Multi2} allow us to present the solution to the ballistic multipole problem in closed form.  This section is devoted to a mathematical discussion of ballistic scattering, while the application of the multipole technique to physical phenomena under current study is deferred to Section~\ref{sec:Photo}--\ref{sec:Atom}.

\subsection{Initial remarks}
\label{sec:Ball1}

As an extension of the method of Green functions $G(\mathbf r,\mathbf r';E)$, the multipole source approach requires knowledge of the latter for a given background potential $U(\mathbf r)$.  However, only for a scarce number of realistic, three-dimensional potentials the Green function is known in closed form.  This selection evidently includes free propagation (Section~\ref{sec:Multi4}), but also comprises the Coulomb potential $U(\mathbf r) = \alpha/|\mathbf r-\mathbf R|$ \cite{Hostler1963a,Hostler1964a}, the isotropic harmonic oscillator \cite{Bakhrakh1971a}, the homogeneous magnetic field \cite{Gountaroulis1972a,Dodonov1975a}, and parallel electric and magnetic fields \cite{Fabrikant1991a,Kramer2001a}.  For the ballistic problem, an analytic expression for the Green function was derived independently by several authors \cite{Dalidchik1976a,Li1990a,Gottlieb1991a}:
\begin{equation}
\label{eq:Ball1.1}
G(\mathbf r,\mathbf r';E) = \frac M{2\hbar^2}  \frac 1{|\mathbf r-\mathbf r'|}
\left[ \Ci(\alpha_+)\Ai\nolimits'(\alpha_-) - \Ci\nolimits'(\alpha_+)\Ai(\alpha_-) \right] \;,
\end{equation}
where the arguments $\alpha_\pm$ of the Airy functions $\Ai(u)$, $\Ci(u) = \Bi(u) + {\rm i}\Ai(u)$ \cite{Abramowitz1965a} are given by:
\begin{equation}
\label{eq:Ball1.2}
\alpha_\pm = - \beta \left[2E + F (z+z') \pm F|\mathbf r - \mathbf r'| \right] \;.
\end{equation}
(A rather elementary derivation of this result is presented in \cite{Bracher1998a}.)  The parameter $\beta$ in (\ref{eq:Ball1.2}) denotes a characteristic inverse energy of the system (and $\beta F$, correspondingly, an inverse length scale) that varies with the force strength $F$:
\begin{equation}
\label{eq:Ball1.3}
\beta = \left( M/4\hbar^2F^2 \right)^{1/3} \;.
\end{equation}

Employing $\beta$, it proves convenient for our calculations to introduce dimensionless quantities for the energy, time, position, and momentum variables:
\begin{eqnarray}
\boldsymbol\rho   = \beta F \mathbf r  \qquad & \epsilon = -2 \beta E  \nonumber \\
\boldsymbol\kappa = \mathbf k/\beta F  \qquad & \tau     = t/2\hbar\beta
\label{eq:Ball1.4}
\end{eqnarray}
(The components of $\boldsymbol\rho$ will be denoted by $\boldsymbol\rho = (\xi,\upsilon,\zeta)^T$.  E.~g., the arguments $\alpha_\pm$ in (\ref{eq:Ball1.1}) read $\alpha_\pm = \epsilon - (\zeta + \zeta') \mp |\boldsymbol\rho -\boldsymbol\rho'|$.)  We use this notation in order to state two helpful integral representations of the Green function (\ref{eq:Ball1.1}).  The first follows from a Laplace transform of the time-dependent ballistic propagator \cite{Kramer2002a}:
\begin{equation}
\label{eq:Ball1.5}
G(\boldsymbol\rho ,\boldsymbol\rho';\epsilon) = - 2{\rm i}\beta (\beta F)^3 \int_0^\infty
\frac{{\rm d}\tau}{({\rm i}\pi\tau)^{3/2}}\, {\rm e}^{{\rm i}(\boldsymbol\rho -\boldsymbol\rho')^2/\tau + {\rm i}\tau (\zeta+\zeta'-\epsilon) -{\rm i}\tau^3/12} \;,
\end{equation}
while the second form features the propagator in momentum space:
\begin{multline}
\label{eq:Ball1.6}
G(\boldsymbol\rho ,\boldsymbol\rho';\epsilon) = - \frac{{\rm i}\beta}{4\pi^3} (\beta F)^3 \int_0^\infty
{\rm d}\tau \, {\rm e}^{-{\rm i}\epsilon\tau - {\rm i}\tau^3/12} \,\times \\
\int {\rm d}^3\kappa \int {\rm d}^3\kappa' \,{\rm e}^{-{\rm i}(\boldsymbol\kappa\cdot\boldsymbol\rho + \boldsymbol\kappa'\cdot\boldsymbol\rho')} \,\delta(\boldsymbol\kappa + \boldsymbol\kappa' + 2\tau\hat e_z)
\, {\rm e}^{-{\rm i}(\boldsymbol\kappa -\boldsymbol\kappa')^2/16} \;.
\end{multline}

Before starting out with the multipole calculations, we remark a general feature of the ballistic problem.  From (\ref{eq:Ball1.1}) and (\ref{eq:Ball1.2}), we infer that $G(\mathbf r,\mathbf r';E)$ is a functional of the variables $\mathbf r-\mathbf r'$ and $E + Fz'$ only.  This translational symmetry is a consequence of the uniformity of the force field, where a shift of the source position merely alters the potential energy of the emitted particles.  We may take advantage of this invariance to relocate the source to the origin \cite{Kramer2002a}:
\begin{equation}
\label{eq:Ball1.7}
G(\mathbf r,\mathbf r';E) = G(\mathbf r-\mathbf r';\mathbf o;E+Fz') \;.
\end{equation}
Furthermore, the exclusive dependence on these variables allows us to replace derivatives of $G(\mathbf r,\mathbf r';E)$ with respect to the source location $\mathbf r'$ by derivatives with respect to $\mathbf r$ and the energy $E$.  One easily verifies that:
\begin{equation}
\label{eq:Ball1.8}
\left(\frac\partial{\partial \xi'}, \frac\partial{\partial \upsilon'}, \frac\partial{\partial \zeta'} \right) G(\boldsymbol\rho, \boldsymbol\rho ';\epsilon) = - \left( \frac\partial{\partial \xi}, \frac\partial{\partial \upsilon}, \frac\partial{\partial \zeta} + 2 \frac\partial{\partial\epsilon} \right) G(\boldsymbol\rho, \boldsymbol\rho ';\epsilon) \;.
\end{equation}
We will summarizingly refer to the exchange of derivatives in (\ref{eq:Ball1.8}) as the substitution rule for the ballistic Green function.

\subsection{Multipole Green functions}
\label{sec:Ball2}

We now characterize the multipole Green functions $G_{lm}(\mathbf r,\mathbf r';E)$ (\ref{eq:Multi2.6}) of the ballistic problem.  In view of the translational symmetry (\ref{eq:Ball1.7}), we remark in advance that it suffices to discuss the case $\mathbf r'\rightarrow \mathbf o$;  the general expression for $G_{lm}(\mathbf r,\mathbf r';E)$ then follows by properly adjusting the positions $\mathbf r$, $\mathbf r'$, and the particle energy $E$.

By definition (\ref{eq:Multi2.6}), the multipole Green function is the spherical tensor gradient $\Klm_{lm}(\boldsymbol\nabla')$ of the ballistic Green function $G(\mathbf r,\mathbf r';E)$ (\ref{eq:Ball1.1}).  Since this differential operator reduces to a polynomial in momentum space, it is advantageous to employ the integral representation (\ref{eq:Ball1.6}) in performing the differentiation.  In the limit $\mathbf r'\rightarrow \mathbf o$, this procedure yields after a suitable shift in the remaining momentum integration variable $\mathbf q = \boldsymbol\kappa + \tau\hat e_z + 2\boldsymbol\rho/\tau$:
\begin{equation}
\label{eq:Ball2.1}
G_{lm}(\boldsymbol\rho,\mathbf o;E) = \frac\beta{4\pi^3} ({\rm i}\beta F)^{l+3} \int_0^\infty {\rm d}\tau\, {\rm e}^{{\rm i}\rho^2/\tau - {\rm i}(\epsilon - \zeta)\tau  -{\rm i}\tau^3/12} \int {\rm d}^3q\, \Klm_{lm} \left( \mathbf q + \tau \hat e_z - 2\boldsymbol\rho/\tau \right) {\rm e}^{-{\rm i}\tau q^2/4} \;.
\end{equation}
The latter integral involves the product of a Gaussian with a polynomial expression and hence allows for evaluation in closed form.  To this end, we first expand the shifted harmonic polynomial $\Klm_{lm} \left( \mathbf q + \tau \hat e_z - 2\boldsymbol\rho/\tau \right)$ into a spherical series with respect to $\mathbf q$, as demonstrated in Appendix~\ref{sec:Translation}.  Since the Gaussian part is isotropic, the only contribution to the integral stems from the term with $\lambda =\mu =0$ in the series (\ref{eq:Trans1}), rendering the calculation trivial.  Furthermore, by a repeated application of the translation theorem for harmonic polynomials (\ref{eq:Trans3}), we may separate the spatial and temporal dependence in the argument of the remaining function $\Klm_{lm}(\tau \hat e_z - 2\boldsymbol\rho/\tau)$.  We then finally obtain a spherical series in $\boldsymbol\rho$ for the momentum integral in (\ref{eq:Ball2.1}):
\begin{equation}
\label{eq:Ball2.2}
\int {\rm d}^3q\, \Klm_{lm} \left( \mathbf q + \tau \hat e_z - 2\boldsymbol\rho/\tau \right) {\rm e}^{-{\rm i}\tau q^2/4} = \frac{8\pi^{3/2}}{({\rm i}\tau)^{3/2}} (-{\rm i})^l \sum_{j = |m|}^l 2^j T_{jlm} ({\rm i}\tau)^{l-2j} \Klm_{jm}(\boldsymbol\rho) \;,
\end{equation}
where the translation coefficients $T_{jlm}$ are given by (\ref{eq:Trans4}).  At this point, it proves convenient to introduce a set of auxiliary functions $\Qk_k(\rho ,\zeta;\epsilon)$ via:
\begin{equation}
\label{eq:Ball2.3}
\Qk_k(\rho,\zeta;\epsilon) = \frac {\rm i}{2\pi\sqrt\pi} \int_0^\infty
\frac{{\rm d}\tau}{({\rm i}\tau)^{k+1/2}}\, \exp\left\{{\rm i}\left[ \frac{\rho^2}\tau+\tau(\zeta-\epsilon)- \frac{\tau^3}{12} \right]\right\} \;.
\end{equation}
Despite their rather involved appearance, for integer $k$ these integrals can be systematically evaluated in closed form, yielding sums over products of Airy functions, as detailed in Appendix~\ref{sec:appendix1}.  (Incidentally, apart from a constant prefactor, the function $\Qk_k(\rho ,\zeta;\epsilon)$ equals the ballistic Green function in $2k+1$ spatial dimensions \cite{Bracher1998a}.  E.~g., a comparison with (\ref{eq:Ball1.5}) reveals that $G(\boldsymbol\rho,\mathbf o;\epsilon) = - 4\beta (\beta F)^3 \Qk_1(\rho ,\zeta;\epsilon)$.)  From (\ref{eq:Ball2.1})--(\ref{eq:Ball2.3}), we then infer that the ballistic multipole Green function $G_{lm}(\mathbf r,\mathbf o;E)$ in the original coordinates is given by:
\begin{equation}
\label{eq:Ball2.4}
G_{lm}(\mathbf r,\mathbf o;E) = -4\beta (\beta F)^{l+3} \sum_{j=|m|}^l (2\beta F)^j T_{jlm} \Klm_{jm}(\mathbf r) \Qk_{2j-l+1}(\beta Fr ,\beta Fz; -2\beta E) \;.
\end{equation}
The explicit expressions obtained from (\ref{eq:Ball2.4}) quickly become cumbersome with increasing multipole order $l$.  While the $s$--wave function $G_{00}(\mathbf r,\mathbf o;E) = G(\mathbf r,\mathbf o;E)/\sqrt{4\pi}$ is displayed already in (\ref{eq:Ball1.1}), here we merely state formulae for the $p$--waves in ballistic scattering \cite{Bracher1999a}:
\begin{multline}
\label{eq:Ball2.5}
G_{10}(\mathbf r,\mathbf o;E) = \sqrt{\frac3\pi} \frac{\beta^3 F^2}{r^3}
\left\{ z \left[ \Ci(\alpha_+)\Ai\nolimits'(\alpha_-) - \Ci\nolimits'(\alpha_+)\Ai(\alpha_-) \right]
\right. \\ \left.
+ 2\beta Fr \left[ \beta \left[ z(2E+Fz) - F r^2 \right]  \Ci(\alpha_+)\Ai(\alpha_-)
+ z \Ci\nolimits'(\alpha_+)\Ai\nolimits'(\alpha_-)  \right]  \right\} \;,
\end{multline}
\begin{multline}
\label{eq:Ball2.6}
G_{1,\pm 1}(\mathbf r,\mathbf o;E) = \sqrt{\frac3{2\pi}} \beta^3 F^2 \frac{x \pm iy}{r^3}
\left\{  \left[ \Ci\nolimits'(\alpha_+)\Ai(\alpha_-) - \Ci(\alpha_+)\Ai\nolimits'(\alpha_-) \right]
\right. \\ \left.
 - 2\beta Fr \left[\Ci\nolimits'(\alpha_+)\Ai\nolimits'(\alpha_-) + \beta \left[2E+Fz\right] \Ci(\alpha_+)\Ai(\alpha_-)  \right]  \right\} \;.
\end{multline}
(The arguments $\alpha_\pm$ of the Airy functions have been defined in (\ref{eq:Ball1.2}).)  The expressions for higher multipole order are, however, quickly calculated by means of (\ref{eq:Ball2.4}).

Despite being exact for all values of the parameters, the complicated structure of the explicit solutions (\ref{eq:Ball2.4}) to the ballistic multipole problem somewhat limits their practical use.  Fortunately, comparatively simple asymptotic approximations for the vicinity of the source as well as for the far-field limit are available from (\ref{eq:Ball2.4}).  We begin with the source region, i.~e., consider the case $\boldsymbol\rho\rightarrow\mathbf o$.  As explained in Section~\ref{sec:Multi3}, in this limit the full multipole orbital symmetry of the corresponding source $\delta_{lm}(\mathbf r')$ (\ref{eq:Multi2.4}) should emerge in the ballistic multipole wave.  This assertion is indeed true:  Replacing the functions $\Qk_{2j-l+1}(\rho,\zeta;\epsilon)$ in (\ref{eq:Ball2.4}) by their principal asymptotic form  for $\boldsymbol\rho \rightarrow \mathbf o$ (\ref{eq:QAsym}), we conclude that the sum term with $j=l$ becomes dominant in the neighbourhood of the source, and with the help of (\ref{eq:Ball1.3}) and (\ref{eq:Trans4}), we finally obtain the leading term in $G_{lm}(\mathbf r,\mathbf o;E)$ as $\mathbf r\rightarrow \mathbf o$:
\begin{equation}
\label{eq:Ball2.7}
G_{lm}(\mathbf r,\mathbf o;E) \sim - \frac M{2\pi\hbar^2}\, (2l-1)!!\, \frac{\Klm_{lm}(\mathbf r)}{r^{2l+1}} \;.
\end{equation}
As expected, the ballistic multipole waves conform to the universally valid expansion (\ref{eq:Multi3.6}) derived in Section~\ref{sec:Multi3}.

In the far-field region $z\rightarrow \infty$, the ballistic multipole waves assume a fairly different character.  Since the presence of the uniform force field manifestly breaks the full rotational symmetry incorporated in (\ref{eq:Multi2.4}), the characteristic angular dependence of the scattering waves propagating in a central potential $U(r)$, as exposed e.~g.\ by the free spherical partial waves (\ref{eq:Multi4.3}), is destroyed.  (Note, however, that the uniform field problem still retains an axis of rotational symmetry.  Consequently, the ballistic Green functions $G_{lm}(\mathbf r,\mathbf o;E)$ (\ref{eq:Ball2.4}) are eigenfunctions of the angular momentum operator $L_z$, and terms with different values of $m$ don't mix.)  For large distances from the source, the external potential bends the scattering waves in the direction of force, and the particle current forms a beam centered around the $z$--axis that continues to spread in the lateral directions while its density profile settles into an invariant shape that reflects the orbital structure of the source.  In a sense, therefore, the force field projects the spatial particle distribution near the source onto an enlarged distant plane image.  (For electrons in a homogeneous force field, this effect has been dubbed ``photodetachment microscopy'' by Blondel et al.\ \cite{Blondel1996a,Blondel1999a}.)  We are thus interested in the asymptotic shape of the ballistic multipole wave in the vicinity of the $z$--axis, i.~e., in the limit $\zeta\rightarrow\infty$ while $\rho -\zeta$ remains finite.  (The exact formula (\ref{eq:Ball2.4}) is of limited use in this region due to large-scale cancellation between the various contributions to the sum.)  Fortunately, there exists a systematic approach (employing a saddle-point approximation) to evaluate all ballistic waves in the far-field sector \cite{Bracher1999a}.  We refrain from displaying it here; instead, we only cite its rather simple result that involves again a harmonic polynomial operator:
\begin{equation}
\label{eq:Ball2.8}
G_{lm}(\mathbf r,\mathbf o;E) \sim -\frac\beta2 (2{\rm i}\beta F)^{l+3} \frac{\Ci(\alpha_+)}{\sqrt{-\alpha_+}} (-1)^m \Klm_{lm} \left( \frac{\beta Fx}{\sqrt{-\alpha_+}}, \frac{\beta Fy}{\sqrt{-\alpha_+}}, {\rm i} \frac\partial{\partial \alpha_-} \right) \Ai(\alpha_-) \;.
\end{equation}
(Here, the argument $\alpha_+$ (\ref{eq:Ball1.2}) approaches $\alpha_+\rightarrow -\infty$, while $\alpha_-$ remains bound.)  Note that all multipole waves share a common propagating factor, whereas the asymptotic beam profile is contained in the derivative of $\Ai(\alpha_-)$.  Explicit expressions for $s$-- and $p$--waves are tabulated in Section~\ref{sec:Photo}.

\subsection{Ballistic multipole currents}
\label{sec:Ball3}

Despite being a fundamental set of quantities, the ballistic multipole Green functions $G_{lm}(\mathbf r,\mathbf o;E)$ (\ref{eq:Ball2.4}) are not directly accessible in experiment.  This privilege is actually reserved for the multipole currents $\mathbf j_{lm,l'm'}(\mathbf r,\mathbf o;E)$ (\ref{eq:Multi2.9}) and $J_{lm,l'm'}(E)$ (\ref{eq:Multi2.12}) which we discuss next.

We start out with a brief analysis of the current density $\mathbf j(\mathbf r)$ (\ref{eq:Multi2.8}) associated to a ballistic scattering wave.  This quantity is accessible from the Green functions $G_{lm}(\mathbf r,\mathbf o;E)$ (\ref{eq:Ball2.4}) through differentiation, and the calculation of the exact matrix elements $\mathbf j_{lm,l'm'}(\mathbf r,\mathbf o;E)$ is tedious yet straightforward.  The resulting expressions are lengthy, and we will refrain from displaying them here except to notice that the current density component $j_{00,00}^{(z)}(\mathbf r,\mathbf o;E)$ in field direction for a ballistic $s$--wave has been published before \cite{Kramer2002a}.  However, the current density profile in the far-field section has been assessed experimentally \cite{Blondel1996a,Blondel1999a}, so it appears worthwhile to deduce asymptotic expressions for the various multipole current matrix elements in this limit.  Since the scattering wave in this regime closely follows the direction of force, we will only examine the $z$ component of the current density vector, i.~e., confine our attention to $j_{lm,l'm'}^{(z)}(\mathbf r,\mathbf o;E)$.  It has been argued \cite{Blondel1999a} that for $z\rightarrow\infty$, the velocity of the particles in the matter wave is determined by their uniform acceleration in the field, regardless of the details of the emission process, and hence is given by $v(z) \sim \sqrt{2Fz/M}$.  In this approximate picture, the resulting current distribution $j^{(z)}(\mathbf r,\mathbf o;E)$ is simply proportional to the particle density $\rho(\mathbf r) = |\psi_{\rm sc}(\mathbf r)|^2$ of the scattering wave:  $j^{(z)}(\mathbf r) \sim \rho(\mathbf r) v(z)$.  This notion is indeed corroborated by a more accurate inspection \cite{Bracher1999a}:  Using the Wronskian relation $\Ci(z)^*\Ci'(z) - \Ci(z)\Ci'(z)^* = -2{\rm i}/\pi$ \cite{Abramowitz1965a}, one quickly arrives from (\ref{eq:Ball2.8}) at the far-field current distribution ($z\rightarrow\infty$):
\begin{multline}
\label{eq:Ball3.1}
j_{lm,l'm'}^{(z)}(\mathbf r,\mathbf o:E) \sim - \frac\beta{4\pi\hbar\alpha_+} (2\beta F)^{l+l'+5} {\rm i}^{l'-l} (-1)^{m+m'} \,\times \\ \left[ \Klm_{lm}\left( \frac{\beta Fx}{\sqrt{-\alpha_+}}, \frac{\beta Fy}{\sqrt{-\alpha_+}}, {\rm i} \frac\partial{\partial\alpha_-} \right) \Ai(\alpha_-) \right]^* \left[ \Klm_{l'm'}\left( \frac{\beta Fx}{\sqrt{-\alpha_+}}, \frac{\beta Fy}{\sqrt{-\alpha_+}}, {\rm i} \frac\partial{\partial\alpha_-} \right) \Ai(\alpha_-) \right] \;.
\end{multline}
Here, $\alpha_-$ (\ref{eq:Ball1.2}) is bound, while $\alpha_+ \rightarrow -\infty$.  The resulting expressions for $l=0,1$ are utilized in Section~\ref{sec:Photo1}.

In the second step, we determine the total current $J(E)$ (\ref{eq:Multi2.10}) carried by a ballistic scattering wave.  Its matrix elements $J_{lm,l'm'}(E)$ (\ref{eq:Multi2.12}) follow from the Green function $G(\mathbf r,\mathbf r';E)$ (\ref{eq:Ball1.1}) by differentiation and a subsequent limiting process, as discussed in Section~\ref{sec:Multi2}.  Before heading towards the calculation of those currents, we note that due to the remaining rotational symmetry of the force field, the total current matrix is diagonal with respect to the quantum number $m$, i.~e., $J_{lm,l'm'}(E) = 0$ for $m\neq m'$.  (See Section~\ref{sec:Multi3}.)

As the multipole formalism invokes spherical tensor gradients $\Klm_{lm}(\boldsymbol\nabla)$ (\ref{eq:Multi2.12}), it is again favorable to perform the actual calculation in the momentum-space representation of the Green function $G(\mathbf r,\mathbf r';E)$ (\ref{eq:Ball1.6}).  The differentiation and limit operations then become trivial, and the matrix elements of the ballistic total current read in integral form:
\begin{multline}
\label{eq:Ball3.5}
J_{lm,l'm'}(E) = \frac\beta{2\pi^3\hbar} (\beta F)^{l+l'+3} \Im\left[ {\rm i}^{l-l'+1} \int_0^\infty {\rm d}\tau \, {\rm e}^{-{\rm i}\epsilon\tau - {\rm i}\tau^3/12} \right. \\ \left. \int{\rm d}^3q\, \Klm_{lm}(\mathbf q-\tau \hat e_z)^* \Klm_{l'm'}(\mathbf q+\tau\hat e_z) \,{\rm e}^{-{\rm i}\tau q^2/4} \right] \;.
\end{multline}
In the next step, the momentum and temporal contributions in the arguments of the harmonic polynomials appearing here are disentangled by means of the translation theorem (\ref{eq:Trans3}) (see Appendix~\ref{sec:Translation}).  This renders the angular integration in (\ref{eq:Ball3.5}) straightforward, and the remaining momentum integral is of Gaussian type, leaving only the temporal integral to be evaluated.  It may be displayed as a sum over the auxiliary functions $\Qi_k(\epsilon)$ closely related to the expression (\ref{eq:Ball2.3}) appearing in the preceding section:
\begin{equation}
\label{eq:Ball3.7}
\Qi_k(\epsilon) = \lim_{\rho,\zeta\rightarrow 0} \Im[\Qk_k(\rho,\zeta;\epsilon )] = \Im \left[ \frac{{\rm i}}{2\pi\sqrt\pi} \int_0^\infty \frac{{\rm d}\tau}{({\rm i}\tau)^{k+1/2}} \, {\rm e}^{-{\rm i}\epsilon\tau - {\rm i}\tau^3/12} \right] \;.
\end{equation}
(A similar set of functions is discussed in \cite{Manakov2000a}.)  Their properties, in particular, their resolution into products of Airy functions, are discussed in Appendix~\ref{sec:appendix2}.  With the definition (\ref{eq:Ball3.7}), the total ballistic multipole currents $J_{lm,l'm'}(E)$ (\ref{eq:Multi2.12}) finally read:
\begin{equation}
\label{eq:Ball3.8}
J_{lm,l'm'}(E) = \delta_{mm'} \frac M{2\pi\hbar^3} (\beta F)^{l+l'+1} (-1)^{l+l'} \sum_{j=|m|}^{\min(l,l')} 2^j (2j+1)!! T_{jlm} T_{jl'm} \Qi_{3j-l-l'+1}(-2\beta E) \;.
\end{equation}
For $l=0,1$, these currents are listed in Section~\ref{sec:Photo2}.

Despite being an exact expression, it is not easy to comprehend the behavior of the total multipole currents from (\ref{eq:Ball3.8}).  To elucidate it, we proceed with an asymptotic expansion of the multipole currents $J_{lm}(E)$ (here, we treat the diagonal matrix elements only) in terms of the energy parameter $\epsilon = -2\beta E$ valid in the sector $|\epsilon| \gg 1$.  Clearly, the problem reduces to finding asymptotic forms for the auxiliary functions $\Qi_k(\epsilon)$ (\ref{eq:Ball3.7}), and its somewhat intriguing solution is displayed in Appendix~\ref{sec:appendix3}.  Here, we merely state the subsequent results for the ballistic current $J_{lm}(E)$.

We have to distinguish between two mathematically and physically rather different cases.  To start out, we consider the limit $\epsilon \gg 1$ pertaining to large negative energies.  This corresponds to the interesting phenomenon of ballistic tunneling from a point source that has no classical counterpart.  The leading asymptotic term for the total multipole current $J_{lm}(E)$ then reads:
\begin{equation}
\label{eq:Ball3.12}
J_{lm}(E) \sim \frac{M\kappa^{2l+1}}{4\pi^2\hbar^3} \frac{(2l+1)(l+|m|)!}{|m|!(l-|m|)!} \left( \frac{\beta F}\kappa \right)^{3|m|+3} {\rm e}^{-\kappa^3/6(\beta F)^3} \;.
\end{equation}
Here, $\kappa = 2\beta F\sqrt{\epsilon}$ denotes the evanescent particle momentum at the source.  As expected, tunneling is exponentially suppressed.  We also note that for fixed $l$, the current strength declines with increasing quantum number $|m|$.  This is evidence for the centrifugal suppression of tunneling discussed below.

In the opposite limit of classical ballistic motion, $k\rightarrow \infty$ (where $k = 2\beta F\sqrt{-\epsilon}$, $\epsilon \ll -1$), the multipole current $J_{lm}(E)$ may be interpreted as a sum of two terms of different character:  The dominant secular contribution is independent of the field strength $F$ and thus obviously must reproduce the total current for freely propagating partial waves (Section~\ref{sec:Multi4}) given by Wigner's law (\ref{eq:Multi4.4}).  The free-particle expression is then modified by an oscillating contribution akin to (\ref{eq:Ball3.12}):
\begin{equation}
\label{eq:Ball3.13}
J_{lm}(E) \sim \frac{M k^{2l+1}}{4\pi^2\hbar^3} \left\{ 1 - (-1)^l \frac{2(2l+1)(l+|m|)!}{|m|!(l-|m|)!} \left( \frac{\beta F}k \right)^{3|m|+3} \cos\left[ \frac16 \left( \frac k{\beta F} \right)^3 + \frac{|m|\pi}2 \right] \right\} \;.
\end{equation}
The modulation becomes most apparent for linear polarization ($m=0$), but even then it is suppressed with respect to the secular part by a factor $k^{-3}$.  However, due to its rapidly oscillating behavior it still may imprint a conspicuous pattern onto the general trend of $J_{lm}(E)$:  For $m=0$, the energy derivative of the current reads asymptotically:
\begin{equation}
\label{eq:Ball3.14}
\frac\partial{\partial E} J_{l0}(E) \sim (2l+1) \frac{M^2 k^{2l-1}}{2\pi^2\hbar^5} \sin\left[ \frac23 (2\beta E)^{3/2} + (-1)^l \frac\pi4 \right]^2 \;.
\end{equation}
Thus, the current $J_{l0}(E)$ grows monotonically with $E$, yet becomes stationary at a series of energy values $E_{\nu l} = [3\pi(4\nu+2l-1)]^{2/3}/8\beta$ ($\nu$ integer, $\nu > -l/2$), rendering a ``staircase'' structure in plots of $J_{l0}(E)$ versus the energy $E$.

\subsection{Semiclassical theory}
\label{sec:Ball4}

In the previous sections, we derived the exact quantum mechanical solution to the ballistic multipole problem as stated in equations (\ref{eq:Ball2.4}), (\ref{eq:Ball3.1}), and (\ref{eq:Ball3.8}), respectively.  Although a major feature of our presentation, the physical content of these mathematical expressions is not extracted easily.  Hence, in the following paragraphs, we will provide some surprisingly simple arguments of essentially semiclassical nature that within their range of validity faithfully reproduce the features of the quantum solution.  As before, we start out with the current distribution in the far-field sector ($z\rightarrow \infty$).

\begin{figure}[t]
\centerline{\includegraphics[draft=false,width=3in]{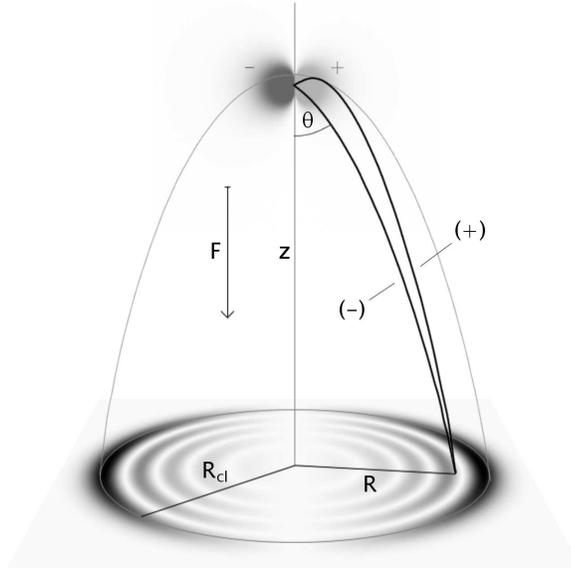}}
\caption{Semiclassical ballistic motion in the far-field limit.  Within the disc $R < R_{\rm cl}$, two parabolic paths (bold) emitted under equal angles $\theta$, $\pi-\theta$ from the multipole source will join the origin with the destination.  Unlike the fast path (+), the slow path (--) undergoes a reflection at the parabolic turning surface $\alpha_-=0$.  The accumulated phases, together with their initial ``atomic'' phases inherited from the point source, determine the exact shape of the interference pattern on the screen.
\label{fig:semicl}}
\end{figure}

Like in the preceding section, we have to distinguish two radically different energy regimes.  In quantum motion, the initial particle energy $E$ can be negative, leading to ballistic tunneling, which is a phenomenon with no immediate classical counterpart.  We will assess it below.  For the moment, however, we consider the simpler case of classically allowed ballistic motion originating from a point source, a problem already studied by Galilei.  Our approach, like some earlier contributions \cite{Fabrikant1981a,Demkov1982a,Du1989b,Kondratovich1990a,Blondel1996a}, is built upon the Hamilton-Jacobi theory of uniformly accelerated motion which is discussed in detail in Ref.~\cite{Bracher1998a}.  We assume that classical bodies of some fixed energy $E$ are emitted at the origin $\mathbf r'=\mathbf o$ and travel towards a destination position $\mathbf r$ on a distant screen perpendicular to the direction of force ($z=\;$const., $z\rightarrow\infty$).  The freely falling bodies then may follow two different parabolic tracks:  The ``fast'' trajectory, denoted by $(-)$, directly leads to the location $\mathbf r$, whereas the ``reflected'' or ``slow'' trajectory $(+)$ first grazes the turning surface, a rotational paraboloid with the source as its focal point given by $\alpha_- = 0$ (\ref{eq:Ball1.2}) which presents the maximum range of classically allowed motion \cite{Bracher1998a}, before arriving at $\mathbf r$.  (The situation is depicted in Figure~\ref{fig:semicl}.)  On the screen the particles will populate the circular disc $\alpha_- < 0$ whose radius $R_{\rm cl}$ is asymptotically given by $R_{\rm cl}^2 = 4Ez/F$.  Correspondingly, as $z\rightarrow \infty$,
\begin{equation}
\label{eq:Ball4.1}
\alpha_- \sim -2\beta E \left( 1 - R^2/R_{\rm cl}^2 \right) \;.
\end{equation}
Thus, for large distances the current profile settles into an invariant shape of absolute size $R_{\rm cl}$.

Next, we discuss the projection properties of the force field for large $z$.  The relative lateral position $R/R_{\rm cl}$ on the screen is determined by the angle of emission $\theta$ relative to the direction of force.  Since motion perpendicular to $\mathbf F$ is uniform, $R$ is the product of the corresponding component of the initial particle velocity with the time of flight:  $R = T(\theta;E) v_i \sin\theta $.  However, in the far-field limit $T(\theta;E)$ is asymptotically independent of $\theta$ and $E$, and we find the simple relationship $R(\theta) = R_{\rm cl}\sin\theta$.  Hence, the cone $\theta=\,$const.\ will be projected onto a circle of radius $R_{\rm cl}\sin\theta$, and trajectories starting under opposite angles $(\theta,\phi)$, $(\pi-\theta,\phi)$ will asymptotically share the same destination $\mathbf r$ on the screen.

The distribution of emission angles at the source and the current profile recorded at the detecting screen are linked through the classical differential cross section $\partial\sigma_{\rm cl}/\partial\Omega$ of the ballistic problem in the asymptotic limit.  For both the direct and reflected path respectively, we find:
\begin{equation}
\label{eq:Ball4.2}
\frac{\partial \sigma_{\rm cl}}{\partial\Omega}\,(R;E) =
\frac{R\,{\rm d}R}{\sin\theta\, {\rm d}\theta} =
R_{\rm cl}^2 \cos\theta = R_{\rm cl}^2 \sqrt{1 - R^2/R_{\rm cl}^2} \;.
\end{equation}
The density of trajectories $\partial\Omega/\partial\sigma_{\rm cl}(R;E)$ diverges at $R_{\rm cl}$, as expected near the caustic surface $\alpha_- = 0$ \cite{Schulman1981a}.  In the actual current density distribution, this divergence will be smoothened out into a bright fringe.  (If, however, emission perpendicular to $\mathbf F$ is suppressed, as e.~g. for a $p_z$ multipole source, the edge of the current spot appears quenched.)

In an entirely classical picture, the differential cross section (\ref{eq:Ball4.2}) would completely determine the particle distribution at large $z$.  However, a fundamental modfication is imposed by the wavelike nature of quanta.  A freely falling particle of fixed energy $E$ will travel along both the ``slow'' and ``fast'' paths, but will accumulate different phases in the process, leading to interference between the particle waves.  As the phase difference depends on the location $\mathbf r$---at $R = R_{\rm cl}$ the classical trajectories will coincide, whereas maximum phase difference occurs for particles emitted (anti-)parallel to the direction of force---constructive and destructive interference will alternately take place, leading to a circular interference pattern superimposed onto the cross section (\ref{eq:Ball4.2}).  Conceptionally, the situation resembles the traditional double-slit setup \cite{Moellenstedt1959a,Feynman1965a} used to discuss two-path interference.  From a practical point of view, the ballistic problem enjoys the advantage that no mechanical ``slit'' is involved as the selection of paths is solely accomplished by the force field.  The predicted interference patterns were first observed experimentally by Blondel et al.\ \cite{Blondel1996a,Blondel1999a}.

A quantitative semiclassical analysis rests on the quantum phases $\sigma_\pm(R,\phi;E)$ carried by the interfering trajectories $(+)$, $(-)$.  Generally, three different terms contribute to $\sigma_\pm(R,\phi;E)$ (see also Figure~\ref{fig:semicl}):  First, the trajectories will ``inherit'' atomic phases $\gamma(\theta,\phi)$, $\gamma(\pi-\theta,\phi)$ deriving from the angular phase distribution of the point source; second, particles traveling along the classical paths will gather dynamical phases which semiclassically are determined by the reduced action $W_{\rm cl}^{(\pm)}(\mathbf r,\mathbf o;E)$ along the respective trajectories \cite{Bracher1998a} which reads in terms of $\alpha_\pm$ (\ref{eq:Ball1.2}):
\begin{equation}
\label{eq:Ball4.3}
W_{\rm cl}^{(\pm)}(\mathbf r,\mathbf o;E) = \frac{2\hbar}3 \left[ (-\alpha_+)^{3/2} \pm (-\alpha_-)^{3/2} \right] \;.
\end{equation}
Finally, the ``slow'' trajectory $(+)$ undergoes an additional ``phase jump'' of $-\pi/2$ due to reflection at the turning surface \cite{Schulman1981a}.  Hence, the semiclassical phases $\sigma_\pm(R,\phi;E)$ along the trajectories $(+)$, $(-)$ read:
\begin{eqnarray}
\label{eq:Ball4.4}
\sigma_+(R,\phi;E) &=& \gamma(\pi-\theta,\phi) + \frac1\hbar\,
W_{\rm cl}^{(+)}(\mathbf r,\mathbf o;E) -\frac\pi2 \;, \\
\label{eq:Ball4.5}
\sigma_-(R,\phi;E) &=& \gamma(\theta,\phi) + \frac1\hbar\,
W_{\rm cl}^{(-)}(\mathbf r,\mathbf o;E) \;.
\end{eqnarray}

We now put together the semiclassical theory of ballistic motion in the limit $z\rightarrow\infty$.  Denoting the angular amplitude distribution at the source by $A(\theta,\phi)=|A(\theta,\phi)| \exp[i\gamma(\theta,\phi)]$, the semiclassical current density distribution $j_z^{\rm (sc)}(R,\phi;E)$ on the screen will be given by the differential cross section $\partial \sigma_{\rm cl}/\partial\Omega (R;E)$ (\ref{eq:Ball4.2}), accounting for the projection properties of ballistic motion, modulated by an oscillating term representing the combined effects of source emission characteristics and quantum interference (where $\sin\theta = R/R_{\rm cl}$):
\begin{equation}
\label{eq:Ball4.6}
j_z^{\rm (sc)}(R,\phi;E) =
\frac{\partial\Omega}{\partial \sigma_{\rm cl}}(R;E)
\left| |A(\pi-\theta,\phi)|{\rm e}^{{\rm i}\sigma_+(R,\phi;E)} +
|A(\theta,\phi)|{\rm e}^{{\rm i}\sigma_-(R,\phi;E)} \right|^2 \;.
\end{equation}

In general, little information about the emission characteristics can be extracted from the current profile.  However, the contributions of source structure and quantum interference will disentangle if the emission pattern shows reflection symmetry with regard to the $x-y$ plane, i.~e., $|A(\pi-\theta,\phi)|^2 = |A(\theta,\phi)|^2$ holds.  Then, (\ref{eq:Ball4.6}) reduces to a product of three independent factors \cite{Demkov1982a}:
\begin{equation}
\label{eq:Ball4.7}
j_z^{\rm (sc)}(R,\phi;E) = 4|A(\theta,\phi)|^2 \frac{\partial\Omega}{\partial \sigma_{\rm cl}}\,(R;E) \cos^2 \left( \frac{\sigma_+(R,\phi;E)-\sigma_-(R,\phi;E)}2 \right) \;.
\end{equation}

This condition always applies for pure multipole sources $\delta_{lm}(\mathbf r)$ (\ref{eq:Multi2.4}).  We now calculate the semiclassical approximation $j_{lm,lm}^{\rm (sc)}(\mathbf r,\mathbf o;E)$ to the ensuing multipole current density distribution $j_{lm,lm}^{(z)}(\mathbf r,\mathbf o;E)$ (\ref{eq:Ball3.1}).  Apart from an overall factor $A_0(E)$ the angular amplitude distribution of this source is given by a spherical harmonic function:  $A(\theta,\phi) = A_0(E)\, Y_{lm}(\theta,\phi)$.  In the classical picture, the total emission rate $J_{lm}(E)$ (\ref{eq:Ball3.8}), and thus $A_0(E)$, remains undefined.  Here, we pragmatically adopt the value quantum theory provides for field-free emission, i.~e., the Wigner law (\ref{eq:Multi4.4}) from Section~\ref{sec:Multi4}, and set $A_0(E)^2 = J_{lm}^{\rm (free)}(E) = Mk^{2l+1}/4\pi^2\hbar^3$.  Inserting an explicit formula \cite{Edmonds1957a} for $Y_{lm}(\theta,\phi)$ into (\ref{eq:Ball4.7}), eliminating the angle $\theta$ and approximating the semiclassical phase $\sigma_\pm(R,\phi;E)$ (\ref{eq:Ball4.3})--(\ref{eq:Ball4.5}) using (\ref{eq:Ball4.2}) we obtain after a short calculation:
\begin{multline}
\label{eq:Ball4.9}
j_{lm,lm}^{\rm (sc)}(\mathbf r,\mathbf o;E) = \frac{M k^{2l+1}}{4\pi^3\hbar^3}
\frac{2l+1}{R_{\rm cl}\sqrt{R_{\rm cl}^2-R^2}} \,\frac{(l-m)!}{(l+m)!}\, P_l^m \left(\sqrt{1-\frac{R^2}{R_{\rm cl}^2}}\right)^2 \,\times\\
\sin\left\{ \frac23 \left[ 2\beta E \left(1-\frac{R^2}{R_{\rm cl}^2}\right)
\right]^{3/2} \pm \frac\pi4 \right\}^2 \;.
\end{multline}
Here, $P_l^m(z)$ denotes the associated Legendre polynomial \index{Legendre polynomials $P_l(x)$}\cite{Abramowitz1965a}.  Multipole sources thus generate two different patterns of sharp ringlike interference fringes.  The upper sign in the interference term applies for even parity, i.~e., for even $l-|m|$, whereas the lower sign is valid for  odd $l-|m|$.  Therefore, under change of parity, the interference ring pattern will reverse.

With increasing energy $E$, new interference fringes spring up in the center of the current distribution.  There, the phase difference $\sigma_+ - \sigma_-$ (\ref{eq:Ball4.4}) is linked to the action functional along the ``(+)'' path segment returning to the source, known as the ``closed orbit'': $W_{\rm co}(E) = 4\hbar(2\beta E)^{3/2}/3$.  It has been noted \cite{Du1989a} that destructive interference in the closed orbit takes place whenever $E=E_{\nu l}$ (see Section~\ref{sec:Ball3}).  Thus, the stationary points in the total current $J_{l0}(E)$ (\ref{eq:Ball3.14}) coincide with the appearance of a new dark interference fringe in the pattern, providing a crude explanation for the staircase structure prominent in the multipole currents with $m=0$.

We now turn to the case of ballistic tunneling ($E<0$).  It is of considerable theoretical interest in being probably the only example of a tunneling process in three spatial dimensions yielding to an analytic solution, and thus forms an important testing ground for semiclassical theories of multi-dimensional tunneling \cite{Kapur1937a,VanHorn1967a,Banks1973a,Huang1990a}.  Unlike their one-dimensional textbook counterpart, realistic tunneling problems are notoriously intricate and show a wealth of surprising features (e.~g., trajectories in complex space) absent in the former.  For a discussion of $s$--wave ballistic tunneling, we refer the reader to Ref.~\cite{Bracher1998a}.

A systematic study of ballistic multipole tunneling is outside the scope of this article \cite{Bracher1999a}.  Rather, we present a simple heuristic approach that might be dubbed ``analytic continuation'':  Even as the notion of a trajectory becomes dubious when applied to tunneling, we formally take over the results (\ref{eq:Ball4.1})--(\ref{eq:Ball4.9}) achieved above for classically allowed motion, subjecting them only to minor modifications.  We note that the square of the classical radius $R_{\rm cl}^2$, as defined by $\alpha_-=0$, is now formally negative.  We replace it by its absolute value $R_{\rm tun}^2 = -R_{\rm cl}^2  = 4|E|z/F$.  Similarly, quantities derived from it will change accordingly;  in particular, we find $\alpha_- \sim -2\beta E(1+R^2/R_{\rm tun}^2)$ (\ref{eq:Ball4.1}), $\partial\sigma_{\rm tun}/\partial\Omega(R,E) = R_{\rm tun}^2\sqrt{1+R^2/R_{\rm tun}^2}$ (\ref{eq:Ball4.2}) for the ``cross section,'' and the ``projection law'' $\cos\theta = \sqrt{1+R^2/R_{\rm tun}^2}$.  Clearly, these expressions are devoid of any geometrical interpretation.  Moreover, the tunneling action functional, and hence the phase along the ``trajectory,'' becomes complex \cite{Bracher1998a}:
\begin{equation}
\label{eq:Ball4.10}
W_{\rm tun}(\mathbf r,\mathbf o;E) = \frac{2\hbar}3 \left[ (-\alpha_+)^{3/2} + {\rm i} \alpha_-^{3/2} \right] \;.
\end{equation}
Since the tunneling wave function must decay exponentially, only one of the two possible extensions of (\ref{eq:Ball4.3}) is physically acceptable.  This restriction lifts the path ambiguity, so interference is absent in the tunneling regime.  Therefore, (\ref{eq:Ball4.6}) is replaced by:
\begin{equation}
\label{eq:Ball4.11}
j_z^{\rm (tun)}(R,\phi;E) = \frac{\partial\Omega}{\partial \sigma_{\rm tun}}(R;E)
|A(\theta,\phi)|^2 \exp\left( -\frac43 \alpha_-^{3/2}\right) \;.
\end{equation}
For the special case of a pure multipole source $\delta_{lm}(\mathbf r)$ (\ref{eq:Multi2.4}), we finally obtain with the evanescent momentum $\hbar\kappa = \sqrt{2M|E|}$, in analogy to (\ref{eq:Ball4.9}):
\begin{multline}
\label{eq:Ball4.12}
j_{lm,lm}^{\rm (tun)}(\mathbf r,\mathbf o;E) = \frac{M\kappa^{2l+1}}{16\pi^3\hbar^3}
\frac{2l+1}{R_{\rm tun}\sqrt{R_{\rm tun}^2+R^2}} \,\frac{(l-m)!}{(l+m)!} \,\left| P_l^m \left(\sqrt{1+\frac{R^2}{R_{\rm tun}^2}}\right)\right|^2  \, \times  \\
\exp\left\{ -\frac43 \left[ -2\beta E \left(1+\frac{R^2}{R_{\rm tun}^2}\right)\right]^{3/2} \right\} \;.
\end{multline}
(For real $z>1$, the Legendre polynomial $P_l^m(z)$ is generally complex and double-valued.  The modulus $|P_l^m(z)|^2$ remains unaffected by this complication, however.)

It is instructive to examine the paraxial limit $R\rightarrow 0$ of (\ref{eq:Ball4.12}).  This yields the following approximation to $j_{lm,lm}^{\rm (tun)}(\mathbf r,\mathbf o;E)$, valid for small lateral distances $R$ \cite{Bracher1998a}:
\begin{equation}
\label{eq:Ball4.13}
j_{lm,lm}^{\rm (tun)}(\mathbf r,\mathbf o;E) \sim \frac{M\kappa^{2l+1}}{16\pi^3\hbar^3} \frac{(2l+1)(l+|m|)!}{2^{2|m|}(|m|!)^2(l-|m|)!} \frac{R^{2|m|}}{R_{\rm tun}^{2|m|+2}} \exp\left( -\frac{\kappa R^2}{2z} - \frac{\kappa^3}{6(\beta F)^3}\right) \;.
\end{equation}
This formula allows for a simple interpretation as a product.  The exponential term $\exp[-\kappa^3/6(\beta F)^3]$ equals the WKB penetration factor for a one-dimensional linear potential ramp, while the prefactor $(R/R_{\rm tun})^{2|m|}$ covers the effect of centrifugal repulsion.  It is worth noting that the tunneling current distribution, apart from this prefactor, possesses approximately Gaussian form:  $J_z^{\rm tun}(R,\phi;E) \propto \exp(-\kappa R^2/2z)$:  Remarkably, unlike the total tunneling current $J_{lm}(E)$, the shape of the lateral current profile is independent of the force strength $F$.  This prediction is experimentally confirmed in field emission from ultrasharp tips \cite{Horch1993a}.  We conclude our investigation by pointing out that integration over the current profile (\ref{eq:Ball4.13}) provides an independent means to calculate the total ballistic multipole current $J_{lm}(E)$ in the tunneling limit.  The result coincides with our earlier finding (\ref{eq:Ball3.12}).

\section{Application to photodetachment processes}
\label{sec:Photo}

The multipole wave functions and currents characterizing quantum ballistic motion transcend merely mathematical interest.  Remarkably, these quantities have recently been assessed experimentally in a rather direct fashion, studying electrons accelerated in a homogeneous electric field and ultracold atoms subject to the gravitational force, respectively.  In this section, we start out with a short description of near-threshold photodetachment of negative ions in a the presence of an external uniform field.  This problem has been studied extensively theoretically (see \cite{Fabrikant1981a,Fabrikant1982a,Demkov1982a,Wong1988a,Du1988a,Du1989a,Du1989b,Kondratovich1990a,Kondratovich1990b,Fabrikant1994a,Golovinskii1997a,Manakov2000a} and references therein) as well as experimentally \cite{Bryant1987a,Gibson1993a,Gibson1993b,Blondel1996a,Blondel1999a,Gibson2001a}.  Being aware that many of the subsequent results have appeared in the literature, we nevertheless chose to discuss the multipole source model of photodetachment as it presents the most coherent, and by far the simplest, description of the effect.  However, we strive to emphasize the less known and probed aspects of the subject.  For a brief introduction to the source formalism for $s$--wave photodetachment in an electric field environment, we refer the reader to our previous article \cite{Kramer2002a}.

\subsection{Photodetachment as a source problem}
\label{sec:Photo0}

In the photodetachment setup, a beam of negatively charged ions traverses the focus of a laser beam whose frequency $\omega$ closely matches the electron affinity of the ion, i.~e.\ the binding energy $E_0$ of the excess charge.  Some ions absorb a laser photon and subsequently emit an electron of energy $E=\hbar\omega - E_0$ into a continuum state to become a neutral atom.  In the presence of a homogeneous electric field $\mathbf F = -e\mathbf E$, these electrons are accelerated towards either a counter, allowing measurement of the total photocurrent $J(E)$ \cite{Gibson1993a,Gibson1993b,Gibson2001a}, or a spatially resolving detector plate that records an image of the photoelectron distribution $j_z(\mathbf r,\mathbf o;E)$ \cite{Blondel1996a,Blondel1999a}.  (For technical reasons, the ion and laser beams and the electric field are usually oriented mutually perpendicular.  We choose the $z$--axis as direction of force.)

In the source formalism (Section~\ref{sec:Multi1}), the photodetachment phenomenon is interpreted as the scattering of the ionic electrons at the quantized electromagnetic laser field $\mathbf A$, i.~e.\ the coupling term $W = -e\hbar^2(\mathbf p\cdot\mathbf A)/Mc$ is treated as the interaction potential.  Assuming that the external electric field and the laser field are sufficiently weak not to disturb the electronic configurations of the initial ionic ground state $|\psi_{\rm ion}\rangle$ and the emerging neutral atom in its ground state $|\psi_{\rm atom}\rangle$, we proceed to project these states, leaving an effective inhomogeneous Schr{\"o}dinger equation for the detached electron (\ref{eq:Multi1.2}):
\begin{equation}
\label{eq:Photo0.1}
\left[ E - \frac1{2M} \left( \mathbf p + \frac ec \mathbf A\right)^2 - U_{\rm atom}(\mathbf r) + Fz \right] \psi(\mathbf r) = \left\langle \psi_{\rm atom} | W | \psi_{\rm ion} \right\rangle \;,
\end{equation}
where $U_{\rm atom}(\mathbf r)$ represents the short-range interaction between the emitted electron and the remaining neutral atom.  In leading approximation, this potential can be neglected.  (In the related photoionization effect \cite{Kondratovich1990a,Nicole2002a}, a long-range Coulomb attraction between the emitted electron and the emerging ion prevails which must be included in the external potential: $U_{\rm ext}(\mathbf r) = Fz + e^2/r$.  The additional term renders the treatment of near-threshold photoionization considerably more difficult.)  In strong electric fields, the effects of $U_{\rm atom}(\mathbf r)$ may become important and must be included in a perturbative rescattering series.  Similarly, the influence of the oscillating laser field on the electronic motion may be taken into account by a Floquet series expansion.  (The electron dynamics in the laser field becomes dominant only for extreme intensities where it causes a wealth of new phenomena, e.~g.\ high-harmonic generation \cite{Gavrila1992a,Lohr1997a}.)  A comprehensive study of these corrections is performed in Ref.~\cite{Manakov2000a}.  However, for moderate fields and laser intensities, and frequencies close to the detachment threshold, they may be safely ignored.

In order to reduce the problem to fit the ballistic multipole formalism discussed in Section~\ref{sec:Ball}, we finally approximate the source term $\sigma(\mathbf r) = \left\langle \psi_{\rm atom} | W | \psi_{\rm ion} \right\rangle$ in (\ref{eq:Photo0.1}) by a properly chosen multipole point source (\ref{eq:Multi2.4}).  To this end, we first note that $\sigma(\mathbf r)$ is limited in extension to the size of the parent ion and thus is much smaller than the initial wavelength of the photoelectron.  Similarly to the case of free-particle sources (Section~\ref{sec:Multi4}), the details of the source structure then may be condensed into a single parameter, the source strength $C$.  (Otherwise, finite-size effects have to be taken into consideration.  For Gaussian sources, the deviation from the point source behaviour has been studied in detail \cite{Kramer2002a}.  See also Section~\ref{sec:Atom}.)  This leaves only the multipole character of the source to be determined, which in turn is fixed by the selection rules for dipole radiation.  Since the emission into channels of higher angular momentum at energies close to threshold is strongly suppressed (as exemplified e.~g.\ by the Wigner law (\ref{eq:Multi4.4}) valid for freely propagating particles), only the lowest permissible multipole order is appreciably populated.  In most cases, this effect leads to isotropic emission of the photoelectron ($l=0$) from the point source \cite{Blondel1996a,Blondel1999a,Gibson2001a,Kramer2002a}.  However, if both the parent ion and the emerging neutral atom possess $S$ ground states, the photoelectron must carry the spin of the absorbed laser photon and is therefore emitted into a $p$--wave, where the distribution onto the various $m$ sublevels is determined by the laser polarization vector $\boldsymbol\epsilon$:
\begin{equation}
\label{eq:Photo0.2}
\left[ E - \frac{p^2}{2M} + Fz \right] \psi(\mathbf r) = C \left( \boldsymbol\epsilon\cdot\boldsymbol\nabla \right) \delta(\mathbf r) = C \sum_{m = -1}^{1} \lambda_{1m} \delta_{1m}(\mathbf r) \;.
\end{equation}
(The most prominent member of this class is the hydrogen ion H$^-$ first studied experimentally by Bryant et~al.\ \cite{Bryant1987a}.)  In this form, the photodetachment problem immediately yields to a description in terms of the ballistic multipole waves $G_{10}(\mathbf r,\mathbf o;E)$ (\ref{eq:Ball2.5}) and $G_{1,\pm1}(\mathbf r,\mathbf o;E)$ (\ref{eq:Ball2.6}).  Some results for the ensuing current distributions are presented below.

\subsection{The far-field current profile}
\label{sec:Photo1}

According to the semiclassical model of quantum ballistic motion (Section~\ref{sec:Ball4}), the uniformly accelerated wave function is characterized by a circular fringe pattern imprinted on the Green function by two-path interference in the force field (Figure~\ref{fig:semicl}), and approximations similar to (\ref{eq:Ball4.9}) were first published by Fabrikant \cite{Fabrikant1981a,Fabrikant1982a}.  Demkov et~al.\ \cite{Demkov1982a} not only improved on these results, but also realized that the interference pattern should be experimentally observable in near-threshold photodetachment.  This assertion was finally verified by Blondel et~al\ \cite{Blondel1996a,Blondel1999a} in a seminal series of experiments:  They recorded the photoelectron distribution generated by various ion beams in a homogeneous electric field with a spatially sensitive detector plate and in the course established a new method for the precise determination of electron affinities (``photodetachment microscopy'').  The observed interference patterns are truly macroscopic; for electrical fields $\mathbf E$ of a few hundred V/m and a source-detector distance $z=0.514\;$m, the fringe diameters are in the mm range.  However, all experiments were performed on ions that detach electrons into $s$--waves.  Due to the low photoabsorption rates imposed by the Wigner law (\ref{eq:Multi4.4}), photocurrent profiles in near-threshold $p$--wave detachment are more difficult to measure and so far have not been recorded.  However, source theory predicts an interesting dependence (\ref{eq:Photo0.2}) of these patterns on the laser polarization $\boldsymbol\epsilon$ that is naturally lacking in isotropic emission.  We discuss some examples below.

First, however, we briefly comment on $s$--wave photodetachment.  In this case, the electron is effectively emitted by an isotropic point source $\sigma(\mathbf r) = C\,\delta_{00}(\mathbf r)$, and its wave function is thus proportional to the ballistic Green function $G(\mathbf r,\mathbf o;E)$ (\ref{eq:Ball1.1}).  The density of electrons collected on the distant detector hence serves as a measure of the current density $j^{(z)}(\mathbf r,\mathbf o;E)$ assigned to the Green function (\ref{eq:Multi1.6}) which is stated in exact form in Ref.~\cite{Kramer2002a}, and is in very good agreement with experiment.  Yet, due to the considerable distance of source and detector, a simplified analysis using the far-field approximations (\ref{eq:Ball2.8}) and (\ref{eq:Ball3.1}) for the ballistic Green function and its current density is equally suited.  For $s$--wave emission from a unit strength source ($C=1$), we obtain:
\begin{equation}
\label{eq:Ball2.9}
G_{00}(\mathbf r,\mathbf o;E) \sim 4{\rm i}\beta (\beta F)^3 \frac{\Ci(\alpha_+)}{\sqrt{-4\pi\alpha_+}} \Ai(\alpha_-) \;,
\end{equation}
(where $\alpha_\pm$ is defined in (\ref{eq:Ball1.2})), while the ensuing current density matrix element reads:
\begin{equation}
\label{eq:Ball3.2}
j_{00,00}^{(z)}(\mathbf r,\mathbf o;E) \sim - \frac{2\beta^6F^5}{\pi^2\hbar\alpha_+} \Ai(\alpha_-)^2 \;,
\end{equation}
Thus, for fixed distance $z$, the electron density varies with the radial distance $R$ like the square of the Airy function $\Ai[-2\beta E(1 - R^2/R_{\rm cl}^2)]^2$ (\ref{eq:Ball4.1}), where $R_{\rm cl}^2 = 4Ez/F$ (see also Section~\ref{sec:Ball4}).  This formula was used by Blondel et al.\ \cite{Blondel1999a} in the analysis of their photodetachment experiment.

Now we turn to the case of $p$--wave photodetachment.  From (\ref{eq:Photo0.2}), we expect that the electron wave function $\psi(\mathbf r)$ is a superposition of the $l=1$ multipole Green functions $G_{10}(\mathbf r,\mathbf o;E)$ (\ref{eq:Ball2.5}) and $G_{1,\pm1}(\mathbf r,\mathbf o;E)$ (\ref{eq:Ball2.6}) whose relative weights depend on the laser polarization $\boldsymbol\epsilon$.  For our purposes, again their far-field asymptotics (\ref{eq:Ball2.8}) suffice:
\begin{equation}
\label{eq:Ball2.10}
G_{10}(\mathbf r,\mathbf o;E) \sim -8\sqrt3 {\rm i}\beta (\beta F)^4 \frac{\Ci(\alpha_+)}{\sqrt{-4\pi\alpha_+}} \Ai\nolimits'(\alpha_-) \;,
\end{equation}
\begin{equation}
\label{eq:Ball2.11}
G_{1,\pm1}(\mathbf r,\mathbf o;E) \sim \mp 4\sqrt3 \beta(\beta F)^5 \frac{\Ci(\alpha_+)}{\sqrt{2\pi}(-\alpha_+)} (x \pm {\rm i}y) \Ai(\alpha_-) \;.
\end{equation}
Their associated current density matrix elements $j^{(z)}_{1m,1m'}(\mathbf r,\mathbf o;E)$ are easily evaluated from eq.~(\ref{eq:Ball3.1}).

Here, we examine four specific setups of the photodetachment experiment in detail.  As noted before, the ion and laser beams, and the direction of force should be in mutually orthogonal orientation; we choose them as the $x$--, $y$--, and $z$--directions of our coordinate system, respectively.  Then, the polarization vector $\boldsymbol\epsilon$ is confined to the $x-z$--plane.  Some natural choices for it are $\boldsymbol\epsilon_\pi = \hat e_z$ (parallel or $\pi$--polarization with respect to the electric field $\mathbf E$), $\boldsymbol\epsilon_\sigma = \hat e_x$ (perpendicular or $\sigma$--orientation), $\boldsymbol\epsilon_{\rm circ} = (\hat e_z + {\rm i}\hat e_x)/\sqrt2$ (circular polarization), and $\boldsymbol\epsilon_{\rm tilt} = (\hat e_z + \hat e_x)/\sqrt2$ (linearly polarized under an angle of $45^\circ$ to $\mathbf E$).  The corresponding multipole sources in (\ref{eq:Photo0.2}) then read, apart from a constant of proportionality:
\begin{eqnarray}
\sigma_\pi(\mathbf r) = \delta_{10}(\mathbf r) \;,\quad &
\sigma_{\rm circ}(\mathbf r) = \frac12 [-{\rm i}\delta_{11}(\mathbf r) + \sqrt2\delta_{10}(\mathbf r) + {\rm i}\delta_{1,-1}(\mathbf r)] \;, \nonumber \\
\sigma_\sigma(\mathbf r) = -\frac1{\sqrt2} [\delta_{11}(\mathbf r) - \delta_{1,-1}(\mathbf r)] \;,\quad &
\sigma_{\rm tilt}(\mathbf r) = \frac12 [-\delta_{11}(\mathbf r) + \sqrt2\delta_{10}(\mathbf r) + \delta_{1,-1}(\mathbf r)] \;.
\label{eq:Photo1.1}
\end{eqnarray}
Note that $\sigma_{\rm circ}(\mathbf r)$ and $\sigma_{\rm tilt}(\mathbf r)$ only differ in the relative phase of their multipole components.  A brief calculation using (\ref{eq:Ball3.1}) then yields the asymptotic photocurrent density profiles generated by these $p$--wave sources:
\begin{equation}
\label{eq:Photo1.2}
j_\pi^{(z)}(\mathbf r,\mathbf o;E) \sim \frac{24\beta^8F^7}{\pi^2\hbar(-\alpha_+)} \Ai'(\alpha_-)^2 \;,
\end{equation}
\begin{equation}
\label{eq:Photo1.3}
j_\sigma^{(z)}(\mathbf r,\mathbf o;E) \sim \frac{24\beta^{10}F^9}{\pi^2\hbar\alpha_+^2} x^2 \Ai(\alpha_-)^2 \;,
\end{equation}
\begin{equation}
\label{eq:Photo1.5}
j_{\rm circ}^{(z)}(\mathbf r,\mathbf o;E) \sim \frac{12\beta^8F^7}{\pi^2\hbar(-\alpha_+)}
\left( \Ai'(\alpha_-) - \frac{\beta Fx}{\sqrt{-\alpha_+}} \Ai(\alpha_-) \right)^2 \;.
\end{equation}
\begin{equation}
\label{eq:Photo1.4}
j_{\rm tilt}^{(z)}(\mathbf r,\mathbf o;E) \sim \frac{12\beta^8F^7}{\pi^2\hbar(-\alpha_+)}
\left( \Ai'(\alpha_-)^2 + \frac{(\beta Fx)^2}{(-\alpha_+)} \Ai(\alpha_-)^2 \right) \;,
\end{equation}
(In a different, rather complicated manner, approximations equivalent to (\ref{eq:Photo1.2}) and (\ref{eq:Photo1.3}) were derived by Golovinskii \cite{Golovinskii1997a}.)

\begin{figure}
\centerline{\includegraphics[draft=false,width=4in]{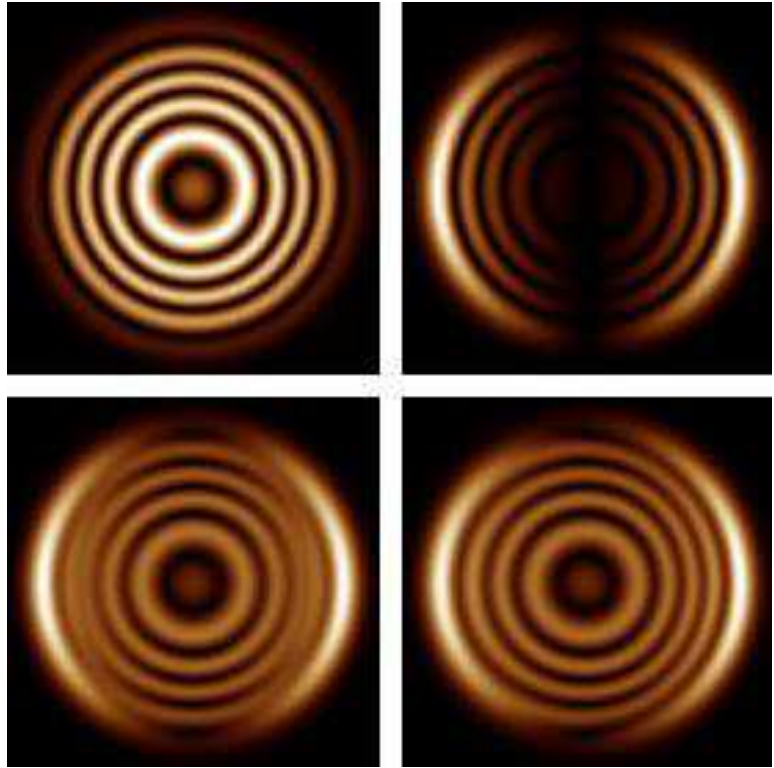}}
\caption{Some photocurrent density profiles in $p$--wave detachment.  Clockwise, from left upper image:  Photoelectron distribution for parallel or $\pi$--polarisation of the laser beam (\ref{eq:Photo1.2}), expected distribution for perpendicular or $\sigma$--polarization (\ref{eq:Photo1.3}), profile for circular polarisation (\ref{eq:Photo1.5}), and linear polarization, tilted under $45^\circ$ from the direction of field (\ref{eq:Photo1.4}).  The area displayed in each image is $1.2\;$mm$\;\times\;1.2\;$mm.  Parameters:  Source-screen distance  $z =0.514\;$m, initial electronic energy $E =60.8\;\mu$eV, electric field strength $\mathbf E = 116\;$V/m.
\label{fig:photodet}}
\end{figure}

We now highlight some properties of these distributions.  They are depicted in Figure~\ref{fig:photodet} for the set of parameters actually used in Blondel's experiment \cite{Blondel1996a}, viz., initial electronic energy $E=6.08\cdot10^{-5}\;$eV, electric field strength $\mathbf E = 116\;$V/m, and detector distance $z=0.514\;$m.  A concentric arrangement of the interference rings only occurs for $\pi$-- and $\sigma$--polarization of the laser beam.  However, the nodes and maxima are interchanged in the respective images.  This effect was already predicted in Section~\ref{sec:Ball4} and may be considered as a hallmark of the differing atomic phase $\gamma(\theta,\phi)$ of the interfering trajectories.  The fact that the fringe pattern encodes information about the orbital structure of the source lead Demkov et~al.\ \cite{Demkov1982a} to its interpretation as an image of the atomic wavefunction, generated by ``photodetachment microscopy.'' (In $\sigma$--polarization, the underlying interference pattern is the same as in $s$--wave photodetachment \cite{Blondel1999a,Kramer2002a}.)  Likewise, a sharply defined interference structure is observed for circular polarization $\boldsymbol\epsilon_{\rm circ}$ (\ref{eq:Photo1.1}), as this source conforms to the condition outlined in (\ref{eq:Ball4.7}).  However, the mirror symmetry $x\rightarrow-x$, present in the other three plots, is conspicuously broken, as the number of fringes differs on the left and right side of the current density profile.  Finally, due to differing weights $|A(\theta,\phi)|$ of the interfering trajectories, no clear-cut ring pattern is present in the tilted case $\boldsymbol\epsilon_{\rm tilt}$.  The corresponding image rather looks like a combination of the patterns obtained in $\pi$-- and $\sigma$--polarization.  Indeed, eq.~(\ref{eq:Photo1.2})--(\ref{eq:Photo1.4}) show that $j_{\rm tilt}^{(z)}(\mathbf r,\mathbf o;E) = [j_\pi^{(z)}(\mathbf r,\mathbf o;E) + j_\sigma^{(z)}(\mathbf r,\mathbf o;E)]/2$ holds.

\subsection{The total photocurrent}
\label{sec:Photo2}

For historical reasons, the total detachment rate $J(E)$ of photoelectrons as a function of the electron excess energy $E$ attracted more attention than the photocurrent distribution $j^{(z)}(\mathbf r,\mathbf o;E)$, despite the fact that $J(E)$ is available from the latter quantity by integration.  Stirred by the experiment of Bryant et~al.\ \cite{Bryant1987a}, a series of theoretical articles approached the problem, generally based either on integral representations of the Green function (\ref{eq:Ball1.1}) \cite{Wong1988a,Fabrikant1994a} or an analysis using Fermi's golden rule (\ref{eq:Multi1.10}) \cite{Du1988a,Du1989a}.  The theory of ballistic multipole source developed in this article yields a closed expression for all multipole photocurrents $J_{lm}(E)$ (\ref{eq:Ball3.8}), thus obliterating the need for the rather involved calculations in the former approaches.

We again briefly comment on the case of isotropic emission of the photocurrent.  For a source of unit strength ($C=1$), we immediately obtain \cite{Kramer2002a}:
\begin{equation}
\label{eq:Ball3.9}
J_{00}(E) = \frac{M\beta F}{2\pi\hbar^3} \left[ \Ai'(\epsilon)^2 - \epsilon\Ai(\epsilon)^2 \right] \;,
\end{equation}
where $\epsilon = -2\beta E$ (\ref{eq:Ball1.4}).  This result was also found by Fabrikant \cite{Fabrikant1994a}, and, via the density of states representation (\ref{eq:Multi1.11}), by Luc--Koenig and Bachelier \cite{LucKoenig1980a}.  The near-threshold $s$--wave photocurrent spectrum has been recorded with high precision in the experiments by Gibson et~al.\ \cite{Gibson1993a,Gibson2001a} and is in virtually perfect agreement with (\ref{eq:Ball3.9}).

For detachment into a $p$--wave, the general formula (\ref{eq:Ball3.8}) reduces to the expressions:
\begin{equation}
\label{eq:Ball3.10}
J_{10}(E) = \frac{M(\beta F)^3}{\pi\hbar^3} \left[ 2\epsilon^2\Ai(\epsilon)^2 - 4\Ai(\epsilon)\Ai'(\epsilon) - 2\epsilon\Ai'(\epsilon)^2 \right] \;,
\end{equation}
\begin{equation}
\label{eq:Ball3.11}
J_{1,\pm1}(E) = \frac{M(\beta F)^3}{\pi\hbar^3} \left[ 2\epsilon^2\Ai(\epsilon)^2 - \Ai(\epsilon)\Ai'(\epsilon) - 2\epsilon\Ai'(\epsilon)^2 \right] \;,
\end{equation}
which differ merely in a single prefactor.  These currents were first obtained via Fermi's golden rule (\ref{eq:Multi1.10}) by Slonim and Dalidchik \cite{Slonim1976a,Manakov2000a}, while integral representations were found by Du and Delos \cite{Du1988a,Du1989a} and Gibson et~al.\ \cite{Gibson1993b}.  Within the experimental error margins, both expressions again reproduce the experimental $p$--wave detachment spectrum \cite{Gibson1993b}.
\begin{figure}[t]
\centerline{\includegraphics[draft=false,width=4in]{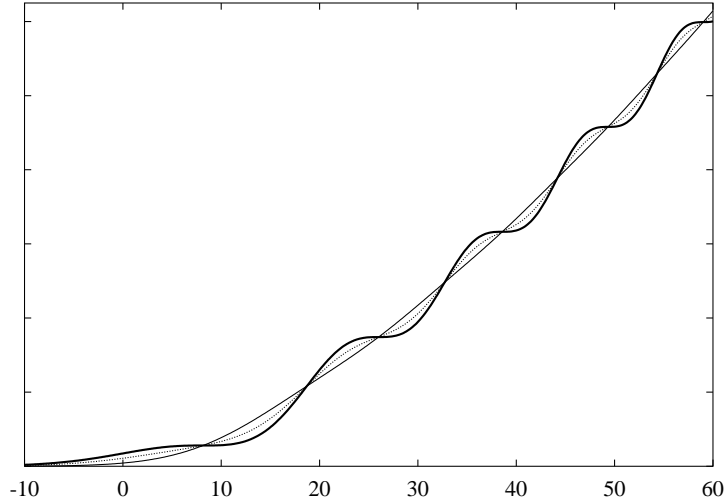}}
\caption{Total current (in arbitrary units) for $p$--wave photodetachment as a function of the electron excess energy $E$, given in $\mu$eV, in an external electric field $\mathbf E = 116\;$V/m.  Bold line:  Current $J_{10}(E)$ (\ref{eq:Ball3.10}), observed in $\pi$--polarisation of the laser beam; light line: $J_{1,\pm1}(E)$ (\ref{eq:Ball3.11}), valid for $\sigma$--polarization.  For both the circular and tilted polarizations (\ref{eq:Photo1.1}), the averaged current is predicted (dotted line).
\label{fig:photocur}}
\end{figure}

Due to the cylindrical symmetry of the potential, the total photocurrent $J(E)$ for arbitrary orientation of the polarization vector $\boldsymbol\epsilon$ can be expressed as a linear combination of (\ref{eq:Ball3.10}) and (\ref{eq:Ball3.11}) (see Section~\ref{sec:Multi2}).  In particular, the current $J_{10}(E)$ is observed in $\pi$--polarization ($\boldsymbol\epsilon\parallel\mathbf E$), while $J_{1,\pm1}(E)$ prevails in $\sigma$--polarization ($\boldsymbol\epsilon\perp\mathbf E$).  In both the tilted and circular polarizations of the laser beam discussed above (\ref{eq:Photo1.1}), the multipole currents contribute equal weight: $J_{\rm circ}(E) = J_{\rm tilt}(E) = [J_{10}(E) + J_{1,\pm1}(E)]/2$.  For illustration, these currents are plotted in Figure~\ref{fig:photocur}.  While generally following their low-field limit given by the Wigner law for $p$--wave emission $J(E) \propto E^{3/2}$ (\ref{eq:Multi4.10}), a strong ``staircase'' modulation is observed in $\pi$--polarization that is absent for the $\sigma$--case $\boldsymbol\epsilon\perp\mathbf E$.  We predicted this behaviour in Section~\ref{sec:Ball3} (\ref{eq:Ball3.14}).

\section{Atom laser with rotating source}
\label{sec:Atom}

In \cite{Kramer2002a} we discuss in detail a model for an atom laser supplied by an ideal BEC.  In this section we will summarize some of the results before we use the presented formalism to extend the theory to higher multipoles.  For the atom laser we consider a two state model, where one magnetically trapped state (the BEC, acting as a quantum source) is weakly coupled to a non-trapped state (the atom laser) by a monochromatic radio-frequency field.  (Replenishing the source condensate to maintain a truly stationary flux in the atom laser poses difficult technical problems.  See \cite{Chikkatur2002a}.)  We are interested in the propagation of coherently released atoms from the condensate in a homogeneous gravitational field.  Explicit closed-form expressions for the wave function of atoms released from an isotropic Gaussian condensate containing $N$ particles:
\begin{equation}
\label{eq:Atom1}
\sigma(\mathbf{r}) = \sqrt N\, \hbar\Omega\,a^{-3/2} \pi^{-3/4} \exp(-r^2/(2a^2)) \;,
\end{equation}
where $\hbar\Omega$ denotes the interaction strength, are given in \cite{Kramer2002a}.  The resulting beam wave function $\psi(\mathbf{r})=\int {\rm d}^3\mathbf{r}' G(\mathbf{r},\mathbf{r'};E)\,\sigma(\mathbf{r'})$ is closely related to the wave function of the point sources discussed in Section~\ref{sec:Ball}.  Similar to extended free-particle sources (\ref{eq:Multi4.8}), in the far-field sector a ballistic Gaussian source may be replaced by a virtual point source at a position displaced from its center along the field $\mathbf F$.  With respect to the virtual point source, we introduce a set of shifted variables depending on the scaled condensate width $\alpha=\beta F a$ (where $\beta$ is defined in (\ref{eq:Ball1.3})), and $\mathbf F = mg\hat e_z$ now represents the gravitational force on a BEC atom:
\begin{equation}
\label{eq:Atom2}
\tilde{\zeta}=\zeta+2\alpha^4 \;,\qquad
\tilde{\rho}^2  =\xi^2+\upsilon^2+\tilde{\zeta}^2 \;,\qquad
\tilde{\epsilon}=\epsilon+4\alpha^4 \;.
\end{equation}
Using these variables, the wave function of the atomic beam $\psi(\mathbf r)$ (\ref{eq:Multi1.4}) and the total current $J_{\rm tot}(E)$ (\ref{eq:Multi1.7}) carried by it are conveniently expressed in terms of the special functions $\Qk_k(\tilde\rho,\tilde\zeta;\tilde\epsilon)$ (\ref{eq:Ball2.3}) and $\Qi_k(\tilde\epsilon)$ (\ref{eq:Ball3.7}) discussed in Appendix~\ref{sec:appendix} (cf.\ \cite{Kramer2002a}):
\begin{equation}
\label{eq:Atom3}
\psi(\mathbf r) = -4\beta{(\beta F)}^3 \Lambda(\tilde{\epsilon}) \Qk_1(\tilde\rho, \tilde\zeta; \tilde\epsilon) \;,
\end{equation}
\begin{equation}
\label{eq:Atom3a}
J_{\rm tot}(E) = \frac8\hbar\, \beta{(\beta F)}^3 \Lambda(\tilde{\epsilon})^2 \Qi_1(\tilde\epsilon) \;.
\end{equation}
We note that the virtual source strength $\Lambda(\tilde\epsilon)$ is strongly energy- and size-dependent:
\begin{equation}
\label{eq:Atom4}
\Lambda(\tilde\epsilon) = \sqrt N\, \hbar\Omega {(2\sqrt{\pi}a)}^{3/2} {\rm e}^{2\alpha^2 \left(\tilde\epsilon - 4\alpha^4/3 \right)} \;.
\end{equation}
In the following we want to use the formalism of Section~\ref{sec:Multi2} to obtain closed solutions for higher states of angular momentum.  These states naturally arise in a rotating condensate.  For simplicity, we assume that the rotating BEC quantum fluid is in its thermodynamical ground state.  This state is known to exhibit a set of vortices (at least, one) symmetrically arranged in an extended lattice structure \cite{Aboshaeer2001a}.

\subsection{Ideal atom laser from a single vortex}
\label{sec:Atom1}

As a first example we describe the atom laser beam arising in the presence of a single vortex with fixed direction in a non-interacting boson gas. The wave function of the rotating condensate is then given by a first excited radial harmonic oscillator state \cite{Fetter2001a}. (In practice, the vortex line is not stationary, but may precess slowly in time \cite{Haljan2001a}.)  For simplicity, we will start with a vortex line aligned to the field direction $\hat e_z$ in an isotropic trap.  In analogy to $p$--wave photodetachment (Section~\ref{sec:Photo}), the source wave function
\begin{equation}
\label{eq:Atom1.1}
\sigma_{11}(\mathbf{r}) = \sqrt N\, \hbar\Omega \,a^{-5/2}\pi^{-3/4}
(x+{\rm i} y)\exp({-r^2/(2a^2)}) \;,
\end{equation}
is the oscillator eigenstate with angular momentum $l=1$ and $m=1$.  The condensate density ${|\sigma_{11}(\mathbf{r})|}^2$ drops to zero along the vortex line.  In analogy to the multipole source formalism (Section~\ref{sec:Multi2}), this wave function may be obtained by applying the (1,1) spherical tensor gradient $\Klm_{11}(\boldsymbol\nabla)$ (\ref{eq:Multi2.2}) to the vortex-free BEC ground state (\ref{eq:Atom1}):
\begin{equation}
\label{eq:Atom1.2}
\sigma_{11}(\mathbf r) = -a\sqrt{\frac{8\pi}{3}} \Klm_{11}(\boldsymbol\nabla)\,\sigma(\mathbf r) \;.
\end{equation}
In general, we define Gaussian multipole sources $\sigma_{lm}(\mathbf r)$ analogous to (\ref{eq:Multi2.4}) via:
\begin{equation}
\label{eq:Atom1.3}
\sigma_{lm}(\mathbf r) = N_l\,\Klm_{lm}(\boldsymbol\nabla)\,\sigma(\mathbf r) \;,
\end{equation}
where $N_l^2 = 2\pi^{3/2}a^{2l}/\Gamma(l+3/2)$ is obtained from the normalization condition $\int {\rm d}^3r \,{|\sigma_{lm}(\mathbf r)|}^2 = N{(\hbar\Omega)}^2$.  In fact, the source functions thus generated turn out to be the lowest-lying oscillator eigenstates of ($l,m$) spherical symmetry (with energy $E = (l+3/2) \hbar^2/Ma^2$).  (In particular, $\sigma_{00}(\mathbf r) = \sigma(\mathbf r)$ covers the isotropic Gaussian source (\ref{eq:Atom1}).)  From (\ref{eq:Multi1.4}) we evaluate the wave function $\psi_{lm}(\mathbf r)$ of the corresponding outcoupled state:
\begin{equation}
\label{eq:Atom1.4}
\psi_{lm}(\mathbf r) = (-1)^l N_l \int{\rm d}^3r' \,\sigma(\mathbf r') \,\Klm_{lm}(\boldsymbol\nabla')\, G(\mathbf r,\mathbf r';E) \;.
\end{equation}
Here, we integrated by parts to shift the spherical tensor operator to the Green function (\ref{eq:Ball1.1}).  The substitution rule (\ref{eq:Ball1.8}) then enables us to further evaluate the integral in terms of derivatives of the known atom laser wave function for zero angular momentum $\psi(\mathbf r)$ (\ref{eq:Atom3}):
\begin{equation}
\label{eq:Atom1.5}
\psi_{lm}(\mathbf r ) = N_l\, \Klm_{lm}\left[ \partial_x, \partial_y, \partial_z-F\partial_E\right] \,\psi(\mathbf r) \;.
\end{equation}
Thanks to the differentiation rules (\ref{eq:QDiffForward}) and (\ref{eq:QDiffBack}), in the far-field sector these derivatives can be expressed as a sum of the auxiliary functions $\Qk_k(\tilde\rho,\tilde\zeta;\tilde\epsilon)$ (\ref{eq:QDef}), analogous to the case of multipole point sources $\delta_{lm}(\mathbf r)$ (\ref{eq:Ball2.4}).  For a single vortex, it suffices to consider the sources $\sigma_{1m}(\mathbf r)$ with $l=1$, $|m|=0,1$:
\begin{eqnarray}
\label{eq:Atom1.6}
\psi_{10}(\boldsymbol\rho) &=& 4\sqrt2\,\beta (\beta F)^3 \alpha \Lambda(\tilde{\epsilon}) \left[
2\tilde{\zeta}\Qk_{2}(\tilde{\rho},\tilde{\zeta};\tilde{\epsilon})
-4\alpha^2    \Qk_{1}(\tilde{\rho},\tilde{\zeta};\tilde{\epsilon})
+             \Qk_{0}(\tilde{\rho},\tilde{\zeta};\tilde{\epsilon})\right] \;,
\\
\label{eq:Atom1.7}
\psi_{1\pm1}(\boldsymbol\rho) &=& \mp 8\beta (\beta F)^3 \alpha  \Lambda(\tilde{\epsilon})
\left(\tilde{\xi}\pm{\rm i}\tilde{\upsilon}\right) \Qk_{2}(\tilde{\rho},\tilde{\zeta};\tilde{\epsilon}) \;,
\end{eqnarray}
(For $\alpha = 0$, point sources are recovered, and (\ref{eq:Atom1.6}), (\ref{eq:Atom1.7}) become proportional to the $p$--wave multipole Green functions (\ref{eq:Ball2.5}), (\ref{eq:Ball2.6}).)  If the vortex is parallel to the gravitational field, the density of the atom laser will be zero along the axis as it is the case for the condensate density:  The vortex line is preserved in the atom laser profile.

In a fashion similar to Section~\ref{sec:Ball3}, we calculate the overall outcoupling rate as a function of the radiation frequency detuning ($E = h\Delta\nu$).  According to (\ref{eq:Multi1.8}), the total multipole current is available from $J_{lm}(E) = -2 \Im [ \langle\sigma_{lm}|G|\sigma_{lm}\rangle ]/\hbar$.  In practice, the calculation is best carried out in momentum space, where both the Gaussian multipole source (\ref{eq:Atom1.3}) and the ballistic propagator (\ref{eq:Ball1.6}) take on a particularly simple form.  Similarly to the point source currents (\ref{eq:Ball3.8}) (to which they reduce as $\alpha \rightarrow 0$), the Gaussian multipole currents are expressed using the auxiliary functions $\Qi_k(\tilde\epsilon)$ (\ref{eq:Ball3.7}) discussed in detail in Appendix~\ref{sec:appendix2}.  Within the $l=1$ triplet, they explicitly read:
\begin{eqnarray}
\label{eq:Atom1.8}
J_{10}(\tilde\epsilon) &=& \frac{32}\hbar\, \beta(\beta F)^3 \alpha^2 \Lambda(\tilde\epsilon)^2
\left[ \Qi_2(\tilde\epsilon)+8\alpha^4\Qi_1(\tilde\epsilon) -4\alpha^2\Qi_0(\tilde\epsilon)+ \frac12\Qi_{-1}(\tilde\epsilon) \right] \;,
\\
\label{eq:Atom1.9}
J_{1\pm1}(\tilde\epsilon) &=&
\frac{32}\hbar\, \beta(\beta F)^3 \alpha^2 \Lambda(\tilde\epsilon)^2 \Qi_2(\tilde\epsilon) \;.
\end{eqnarray}
Thanks to the preserved rotational symmetry of the system, all total current matrix elements $J_{lm,l'm'}(\tilde\epsilon)$ with $m \neq m'$ vanish, as indicated in Section~\ref{sec:Multi2}.

If $\tilde\epsilon$ is large, i.~e., for extended condensates with $\alpha \gg 1$, it is sufficient to use the asymptotic exponential form of the functions $\Qi_k(\tilde\epsilon)$ (\ref{eq:Qi+large}).  Further expanding the currents around their maximum near $\epsilon=0$, we obtain their large-source approximations:
\begin{eqnarray}
\label{eq:Atom1.10a}
J_{10}(\tilde\epsilon) &\sim& N \sqrt\pi\,\beta\hbar\Omega^2 \frac{\epsilon^2}{\alpha^3} \,{\rm e}^{- \epsilon^2/4\alpha^2} \;,
\\
\label{eq:Atom1.10b}
J_{1\pm1}(\tilde\epsilon) &\sim& 2N \sqrt\pi\,\hbar\Omega^2 \frac\beta\alpha \,{\rm e}^{- \epsilon^2/4\alpha^2} \;.
\end{eqnarray}
As expected from our earlier results for a simple Gaussian source \cite{Kramer2002a}, these currents can be interpreted as the integrated condensate density along a slice through the BEC at a height $z$ fixed by the ``resonance condition'' $E+Fz = 0$ (Franck--Condon principle):
\begin{equation}
\label{eq:Atom1.11}
J_{lm}(E) \sim \frac{2\pi}{\hbar} \int{\rm d}^3\mathbf r\, {|\sigma_{lm}(\mathbf r)|}^2\,\delta(E+F z) \;.
\end{equation}
(A semiclassical derivation of the ``slicing approximation'' is presented in Ref.~\cite{Japha2002a}.)  We note that (\ref{eq:Atom1.11}) evidently fulfils the sum rule (\ref{eq:Multi1.9}) for the total outcoupling rate.

\begin{figure}
\centerline{\includegraphics[draft=false,width=4in]{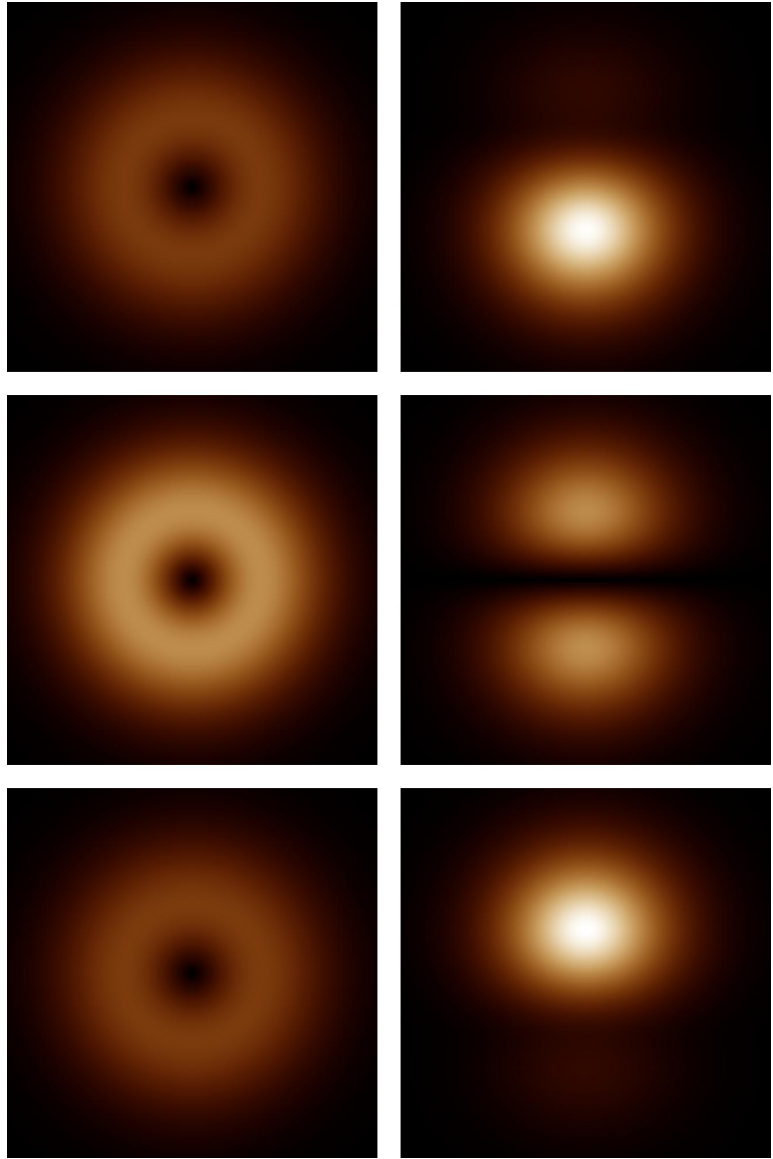}}
\caption{Atom laser density profiles for a rotating $^{87}$Rb BEC source sustaining one vortex.  Left column: Parallel orientation of vortex line and force $\mathbf F$, with source $\sigma_{11}(\mathbf{r})$ (\ref{eq:Atom1.1}).  Right column:  Vortex perpendicular to $\mathbf F$, as given by $\sigma_{1\perp}(\mathbf r)$ (\ref{eq:Atom1.12}).  The detuning frequencies $\Delta\nu$ are $-4\;$kHz (top row), $0\;$kHz (center row), and $+4\;$kHz (bottom row), respectively.  The brightest spots of the distribution pertain to a density of $2.5\;$atoms$/\mu$m$^3$.  Displayed area: $30\,\mu$m $\times$ $30\,\mu$m, distance from source: $z=1\;$mm; source parameters: $a=2\;\mu$m, $\Omega=2\pi \times 100\;$Hz, $N = 10^6\;$atoms.
\label{fig:vortex1_psi}}
\end{figure}

For illustration, we consider two orientations of the vortex with respect to the gravitational force $\mathbf F$.  A vortex parallel to the field is simply represented by the Gaussian condensate wave function $\sigma_{11}(\mathbf r)$ (\ref{eq:Atom1.1}), and the ensuing laser beam characteristics are expressed in (\ref{eq:Atom1.7}) and (\ref{eq:Atom1.9}).  We also examine the case of a vortex along the $x$--axis, i.~e., perpendicular to $\mathbf F$.  The corresponding BEC wave function $\sigma_{1\perp}(\mathbf r)$ is connected to the parallel vortex model by a rotation $\exp(-{\rm i}\pi\hat L_y/2)$.  Application of the $l=1$ rotation matrix for angular momentum eigenstates \cite{Edmonds1957a} yields the following source term:
\begin{equation}
\label{eq:Atom1.12}
\sigma_{1\perp}(\mathbf r) = \frac12 \left[ \sigma_{11}(\mathbf r) + \sqrt2\,\sigma_{10}(\mathbf r) + \sigma_{1,-1}(\mathbf r) \right] \;,
\end{equation}
with the associated total current:
\begin{equation}
\label{eq:Atom1.13}
J_{1\perp}(\tilde\epsilon) = \frac12 \left[ J_{11}(\tilde\epsilon) + J_{10}(\tilde\epsilon)\right] \;.
\end{equation}
Figure~\ref{fig:vortex1_psi} depicts atom laser density profiles generated by an ideal $^{87}$Rb BEC of width $a = 2\;\mu$m at a distance $z =1\;$mm in the center of the resonance ($\nu=0$) as well as for positive and negative detuning ($\nu=\pm4\;$kHz).  For this choice of parameters, $\alpha \approx 3.33$, so the effective energy $\tilde\epsilon = 4\alpha^4$ (\ref{eq:Atom2}) of the assigned virtual point source indicates tunneling emission.  According to the results presented in Section~\ref{sec:Ball4}, we expect the density profiles to be of overall Gaussian shape, with a mean width $D(z) = 2\tilde z/\tilde\kappa$ (\ref{eq:Ball4.13}), where $\tilde z = (\zeta + 2\alpha^4)/\beta F$ and $\tilde\kappa = 4\beta F\alpha^2$ denote the distance from the virtual source and the virtual evanescent wave number, respectively.  (Alternatively, this result follows from the spreading of a minimal uncertainty wave packet of width $a$ during its time of flight $T=\sqrt{2Mz/F}$ \cite{Bracher1998a}.)  This Gaussian envelope is modulated by a factor $f(\xi,\upsilon)$ that depends on the relative orientation of the vortex and the gravitational force.  Indeed, a fairly cumbersome calculation yields for the asymptotic shape of the density profiles generated by the sources (\ref{eq:Atom1.1}) and (\ref{eq:Atom1.12}) in the far-field sector, valid for $\alpha \gg 1$:
\begin{equation}
\label{eq:Atom1.14}
\rho(\xi,\upsilon) \sim 16 N (\hbar\Omega)^2 \beta^5 F^3 \alpha^3 \frac{f(\xi,\upsilon)}{\sqrt{2\pi\zeta} (\zeta + 2\alpha^4)^2} \exp\left[- \left( \frac{\epsilon^2}{4\alpha^2} + \frac{2\alpha ^2(\xi^2 + \upsilon^2)}{\zeta + 2\alpha^4} \right)\right] \;,
\end{equation}
where the modulation factors $f_{11}(\xi,\upsilon)$ and $f_{1\perp}(\xi,\upsilon)$ for parallel and perpendicular orientation read, respectively:
\begin{equation}
\label{eq:Atom1.15}
f_{11}(\xi,\upsilon) = \xi^2 + \upsilon^2 \;,\quad
f_{1\perp}(\xi,\upsilon) = \frac{\epsilon^2}4 + \left( \upsilon - \frac{\epsilon\sqrt\zeta}{2\sqrt2\,\alpha^2} \right)^2 \;.
\end{equation}
(We used the dimensionless coordinates introduced in (\ref{eq:Ball1.4}).)  Clearly, $f_{11}(\xi,\upsilon)$ effects the propagation of the vortex in the parallel case.  However, the dependence of $f_{1\perp}(\xi,\upsilon)$ on the source distance $\zeta$ and the detuning $\nu = -\epsilon/2\hbar\beta$ renders the atom laser profiles generated in perpendicular orientation more intriguing (see Figure~\ref{fig:vortex1_psi}):  The detuning-dependent, isotropic contribution $\epsilon^2/4$ competes with a shifted parabolic term that grows linearly with the detector distance $\zeta$.  For $\zeta \ll 2\alpha^4$, detuning blurs the simple vortex image present at center resonance, while for $\zeta \gg 2\alpha^4$ the latter term in (\ref{eq:Atom1.15}) dominates, causing the appearance of a node line in the profile whose relative position shifts linearly with the detuning $\nu$.  The transition between these markedly different regimes of the atom laser occurs at considerable distance from the BEC.  In our example, we find $\zeta=2\alpha^4$ for a separation $z \approx 150\,\mu$m, and the figure depicts the far-field behaviour.  We note, however, that this characteristic distance grows with the fourth power of the source size $a$ and quickly reaches macroscopic dimensions:  For an atom laser supplied by a BEC of width $a=10\,\mu$m, we find $z\approx10\;$cm!

\begin{figure}
\centerline{\includegraphics[draft=false,width=4in]{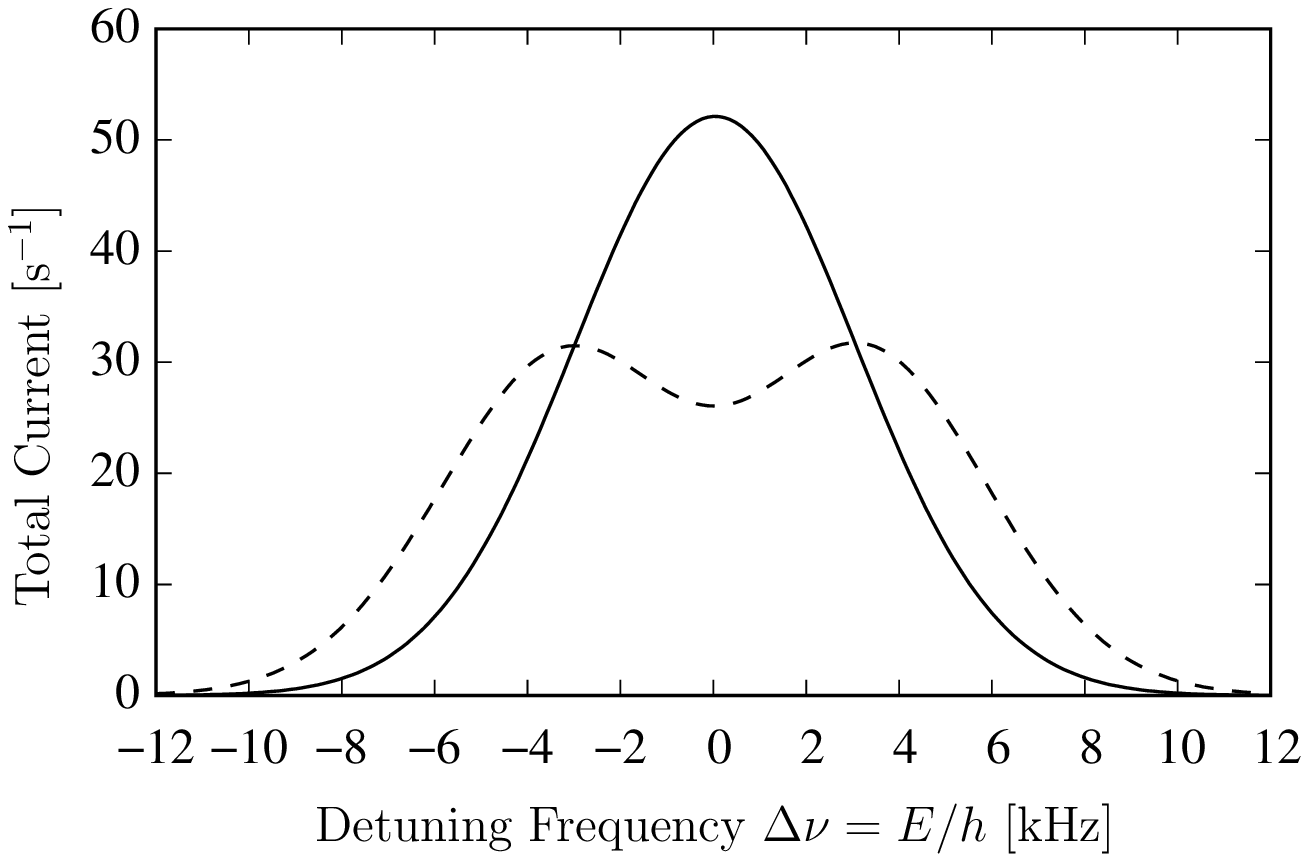}}
\caption{Total current generated per atom in a $^{87}$Rb BEC with one vortex as a function of the detuning frequency $\Delta\nu$. Solid line: The source $\sigma_{11}(\mathbf{r})$ (\ref{eq:Atom1.1}) with the vortex line parallel to the gravitational field $\mathbf F$.  Dashed line: Vortex line perpendicular to $\mathbf F$, represented by $\sigma_{1\perp}(
\mathbf r)$ (\ref{eq:Atom1.12}).  Parameters: $a=2\;\mu$m, $\Omega=2\pi \times 100\;$Hz.
\label{fig:vortex1_jtot}}
\end{figure}

We now turn to the frequency dependence of the total outcoupling rate.  Since $\alpha \gg 1$, use of the asymptotic description (\ref{eq:Atom1.11}) is in order.  In Figure~\ref{fig:vortex1_jtot} we show the resulting current characteristics for both condensate orientations.  While in parallel orientation the current distribution is simply Gaussian, it features a dip in the total current at $\Delta\nu = 0$ for a vortex line perpendicular to the gravitational field.  This behaviour is easily understood from (\ref{eq:Atom1.11}):  Due to the presence of the vortex line in the slicing plane, at $z=0$ the condensate density adopts a minimum.  We finally note here that the slicing approximation fails for small condensates (with $\alpha\sim 1$) and it becomes necessary to use the exact results (\ref{eq:Atom1.8}) and (\ref{eq:Atom1.9}).  The transition between both regimes is studied in detail in \cite{Kramer2002a}.

\subsection{Vortex lattices}
\label{sec:Atom2}

Using the tools developed in the previous section, we proceed to give an example of a multipole source where larger values of the angular momentum are present.  Rotating Bose-Einstein condensates show superfluid behaviour and respond to an externally induced rotation by the formation of a vortex lattice \cite{Matthews1999b,Madison2000a,Aboshaeer2001a,Engels2002a}.  (A review on vortices in Bose-Einstein condensates is given in \cite{Fetter2001a}.)  No attempt at a description of the formation and parameters of this lattice will be made.  Rather, we present a theoretical model for a stationary atom laser supplied by an already formed vortex lattice, with vortex lines oriented parallel to the gravitational field $\mathbf F$.  The wave function of the lattice state (the laser source) is most conveniently modeled in the rotating reference frame (rf), where it becomes time-independent; we denote it by $\sigma_{\rm latt,rf}(\mathbf{r})$.  However, the laser is observed in the laboratory frame and therefore we first have to discuss the transformation between both frames.

\subsubsection{Transformation to the laboratory frame}

In the laboratory frame (lf), the stationary source function for the rotating
frame $\sigma_{\rm latt,rf}(\mathbf{r})$ becomes explicitly time dependent.  The transformation between both frames of reference involves a uniform rotation around the $z$--axis.  This rotation is generated by the unitary operator $\exp\left(-{\rm i}L_z\Omega_{\rm rot} t/\hbar\right)$, where $L_z$ is the $z$ component of the angular momentum operator and $\Omega_{\rm rot}$ is the frequency of rotation. The full time-dependent source term in the lab frame consequently reads:
\begin{equation}
\label{eq:SigmaLatticeLF}
\sigma_{\rm latt,lf}(\mathbf{r},t)=\exp(-{\rm i} E t/\hbar)\exp\left(-{\rm i}L_z\Omega_{\rm rot} t/\hbar\right)\sigma_{\rm latt,rf}(\mathbf{r}) \;.
\end{equation}
The laser wave function resulting from a time-dependent source in the presence of the gravitational field is generated by the ballistic propagator (\ref{eq:Ball1.5}):
\begin{equation}
\label{eq:Atom2.2}
\psi_{\rm latt,lf}(\mathbf{r},t)=-
\frac{{\rm i}}{\hbar} \int {\rm d}^3\mathbf{r'}
\int_{-\infty}^{t}{\rm d}t'\,
K(\mathbf{r},t|\mathbf{r}',t')\,\sigma_{\rm latt,lf}(\mathbf{r}',t') \;.
\end{equation}
Introducing the time of flight $T=t-t'$, and noting that the generator of the rotation $L_z$ commutes with the Hamiltonian of the accelerated motion $H_{\text{grav}}={\mathbf p}^2/2M-Mgz$, this may be rewritten:
\begin{equation}
\label{eq:Atom2.3}
\psi_{\rm latt,lf}(\mathbf r,t) =
-\frac{{\rm i}}{\hbar} {\rm e}^{-{\rm i} (E+L_z\Omega_{\rm rot})t/\hbar}
\int_{0}^{\infty}{\rm d}T\, {\rm e}^{{\rm i} (E + L_z\Omega_{\rm rot}) T/\hbar}\,
\int {\rm d}^3\mathbf{r'} \,K(\mathbf r,T|\mathbf r',0) \sigma_{\rm latt,rf}(\mathbf{r}') \;.
\end{equation}
Thus, the outcoupled laser wave function $\psi_{\rm latt,lf}(\mathbf{r},t)$ is again stationary in the rotating frame:  Like the BEC, the atomic beam profile rotates uniformly with frequency $\Omega_{\rm rot}$.

In the next step, we decompose $\sigma_{\rm latt,rf}(\mathbf{r})$ into a superposition of angular momentum eigenstates:
\begin{equation}
\label{eq:Atom2.4}
\sigma_{\rm latt,rf}(\mathbf{r}) = \sum_{m} \sigma_{m,\rm rf}(\mathbf{r}) \;,
\end{equation}
where $L_z\,\sigma_{m,\rm rf}(\mathbf{r})=m\hbar\,\sigma_{m,\rm rf}(\mathbf{r})$.  In the laboratory frame, the rotating source function appears split into magnetic sublevels shifted in energy:
\begin{equation}
\label{eq:Atom2.5}
\sigma_{\rm latt,lf}(\mathbf{r},t) = {\rm e}^{-{\rm i} E t/\hbar} \sum_{m}
{\rm e}^{-{\rm i} m \Omega_{\rm rot} t} \sigma_{m, \rm rf}(\mathbf{r}) \;.
\end{equation}
Employing the propagator representation of the ballistic Green function (\ref{eq:Ball1.5}), we obtain for the beam wave function:
\begin{equation}
\label{eq:PsiLatticeLF}
\psi_{\rm latt,lf}(\mathbf r,t) =
\sum_{m}{\rm e}^{-{\rm i} (E+m\hbar\Omega_{\rm rot}) t/\hbar}
\int {\rm d}^3\mathbf r'\, G(\mathbf r,\mathbf r';E+m\hbar\Omega_{\rm rot})\,
\sigma_{m, \rm rf}(\mathbf r') \;.
\end{equation}
The rotating source function thus yields to a description in terms of stationary sources, and the results of Section~\ref{sec:Ball} apply.

\subsubsection{The source function in the rotating frame}

The vortex state of the BEC is commonly described as a superposition of angular momentum eigenstates of the harmonic oscillator \cite{Butts1999a,Ho2001a}.  The number of vortices and their positions are available from minimizing the energy functional in the rotating frame.   For a parallel arrangement of vortices and field $\mathbf F$, we may model the vortex state as a product of a two-dimensional ``lattice function'' $\sigma_{2D}(x,y)$ detailing the vortex positions ($x_k,y_k$) with a Gaussian envelope enforced by the harmonic trap potential.  Introducing complex coefficients $v_k=x_k+{\rm i}y_k$, the lattice function is obtained as a product involving all vortex positions that alternatively may be expressed as a polynomial in $(x+{\rm i}y)$,
\begin{equation}
\label{eq:Atom2.7}
\sigma_{2D}(x,y) = \prod_{k=1}^{n}\left[(x+{\rm i}y)-v_k\right] =
\sum_{k=0}^{n}w_k^{(n)}(x+{\rm i}y)^k \;,
\end{equation}
where the coefficients $w_k^{(n)}$ are connected to the vortex positions $v_k$ by the recursion relation $w_k^{(n)} = w_{k-1}^{(n-1)} - v_{k+1} w_k^{(n-1)}$, where $w_{0}^{(0)}=1$.  (Usually, these lattices show rotational symmetry which enforces selection rules on the $w_k^{(n)}$, leaving only few nonvanishing coefficients.)  The complete three-dimensional source function in the rotating frame then reads:
\begin{equation}
\label{eq:Atom2.8}
\sigma_{\rm latt,rf}(\mathbf r) = N_n\exp\left(-\frac{x^2+y^2}{2a_x^2}-\frac{z^2}{2a_z^2}\right)\,\sigma_{2D}(x,y) \;.
\end{equation}
The constant $N_n$ is determined by the normalization condition $\int{\rm d}^3\mathbf r\,|\sigma(\mathbf r)_{\rm latt,rf}|^2 = N(\hbar\Omega)^2$:
\begin{equation}
\label{eq:Atom2.9}
N_n = \frac{\sqrt N\, \hbar\Omega}{\pi^{3/4} \sqrt{a_z \sum_{k=0}^{n} k! |w_k^{(n)}|^2 a_x^{2k+2}}} \;.
\end{equation}
Equation~(\ref{eq:Atom2.7}) then yields the decomposition of $\sigma_{\rm latt,rf}(\mathbf r)$ into eigenstates $\sigma_{m,\rm rf}(\mathbf r)$ of $L_z$ (\ref{eq:Atom2.4}):
\begin{equation}
\label{eq:Atom2.10}
\sigma_{m,\rm rf}(\mathbf r) = N_n w_m^{(n)} {(x+{\rm i}y)}^m \exp\left(-\frac{x^2+y^2}{2a_x^2}-\frac{z^2}{2a_z^2}\right) \;.
\end{equation}
Thus, the $n+1$ source components are all eigenstates of the harmonic trap potential, and the highest quantum number $m$ equals the number of vortices present in the BEC.

Here, we examine the special case of an isotropic trap ($a_x=a_z=a$), for which the theory outlined in the preceding section will deliver the outcoupling rate as well as the beam profile in analytic form.  To this end, we note that the source components $\sigma_{m,\rm rf}(\mathbf r)$ (\ref{eq:Atom2.10}) then simultaneously present eigenstates of $L_z$ and the total angular momentum $L^2$ with quantum number $l=m$.  Hence, the source is entirely made up from circular Gaussian multipole states $\sigma_{mm}(\mathbf r) = N_m \Klm_{mm}(\boldsymbol\nabla) \sigma(\mathbf r)$ (\ref{eq:Atom1}), (\ref{eq:Atom1.3}):
\begin{equation}
\label{eq:Atom2.11}
\sigma_{m,\rm rf}(\mathbf r) =
\sqrt{\frac{m!}{\sum_{k=0}^n k! |w_k^{(n)}|^2 a^{2k}}}\, w_m^{(n)} a^m \sigma_{mm}(\mathbf r) \;.
\end{equation}
According to (\ref{eq:PsiLatticeLF}), the rotating beam is thus produced by a weighed superposition of the stationary sources $\sigma_{mm}(\mathbf r)$, where the effective energy $E_m = E+m\hbar\Omega_{\rm rot}$ of the various components depends on their quantum number $m$.  As explained in the introduction to Section~\ref{sec:Atom}, outside the source region each Gaussian multipole source $\sigma_{lm}(\mathbf r)$ (\ref{eq:Atom1.3}) may be mapped onto a corresponding displaced virtual point source of adjusted strength $\Lambda(\tilde\epsilon)$.  This allows to calculate the wave function $\psi_{mm}(\mathbf r)$ generated by $\sigma_{mm}(\mathbf r)$ along the lines presented in Section~\ref{sec:Ball2}, and the final result closely resembles the corresponding ballistic multipole Green function $G_{mm}(\mathbf r,\mathbf o; E)$ (\ref{eq:Ball2.4}):
\begin{equation}
\label{eq:Atom2.12}
\psi_{mm}(\mathbf r) = -4\beta (\beta F)^3 \Lambda(\tilde\epsilon_m) \,\frac1{\sqrt{m!}} \left[ 2\alpha \left( \xi+{\rm i}\upsilon \right)\right]^m \Qk_{m+1}(\tilde\rho, \tilde\zeta; \tilde\epsilon_m) \;,
\end{equation}
where $\tilde\epsilon_m = -2\beta E_m + 4\alpha^4$ (\ref{eq:Atom2}).  Similarly, the total current $J_{mm}(E_m)$ generated by $\sigma_{mm}(\mathbf r)$ is available from a calculation in the spirit of Section~\ref{sec:Ball3}:
\begin{equation}
\label{eq:Atom2.13}
J_{mm}(E_m) = \frac8\hbar\, \beta (\beta F)^3 (2\alpha)^{2m} \Lambda(\tilde\epsilon_m)^2 \Qi_{m+1}(\tilde\epsilon_m) \;.
\end{equation}
(For $m=0,1$, these expressions reduce to the results (\ref{eq:Atom3})--(\ref{eq:Atom3a}) and (\ref{eq:Atom1.7}), (\ref{eq:Atom1.9}) presented above.)  Substituting (\ref{eq:Atom2.11}) and (\ref{eq:Atom2.12}) in (\ref{eq:PsiLatticeLF}), the wave function of the rotating atom laser beam ultimately reads:
\begin{equation}
\label{eq:Atom2.14}
\psi_{\rm latt,lf}(\mathbf r,t) = \sum_{m=0}^n {\rm e}^{-{\rm i} (E+m\hbar\Omega_{\rm rot}) t/\hbar}
\frac{\sqrt{m!}\, w_m^{(n)} a^m}{\sqrt{\sum_{k=0}^n k! |w_k^{(n)}|^2 a^{2k}}} \,\psi_{mm}(\mathbf r) \;.
\end{equation}
Since cylindrical symmetry enforces that all off-diagonal elements of the total current matrix $J_{lm,l'm'}(E)$ (\ref{eq:Multi2.12}) vanish (see Section~\ref{sec:Multi2}), the (stationary) outcoupling rate $J_{\rm latt}(E)$ reduces to a properly weighed sum of the ballistic multipole currents $J_{mm}(E_m)$ (\ref{eq:Atom2.13}):
\begin{equation}
\label{eq:Atom2.15}
J_{\rm latt}(E) = \frac1{\sum_{k=0}^n k! |w_k^{(n)}|^2 a^{2k}} \sum_{m=0}^n
m!|w_m^{(n)}|^2 a^{2m} \,J_{mm}(E_m) \;.
\end{equation}

\begin{figure}
\centerline{\includegraphics[draft=false,width=4in]{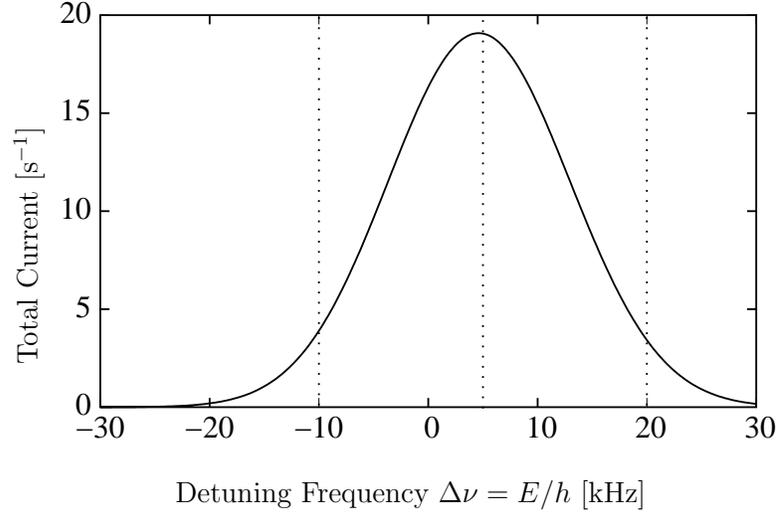}}
\caption{Total current per atom in a $^{87}$Rb BEC with 37 vortices as a function of detuning frequency $\Delta\nu$. The vortex structure of the BEC is not visible in the integrated current. Vortex separation in the BEC: $10\;\mu$m, rotation frequency $\Omega_{\rm rot}=2\pi\times 250\;$Hz, outcoupling strength $\Omega = 2\pi \times 100\;$Hz, size parameter $a = 5\;\mu$m.
\label{fig:vortex37_jtot}}
\end{figure}

\begin{figure}
\centerline{\includegraphics[draft=false,width=5.2in]{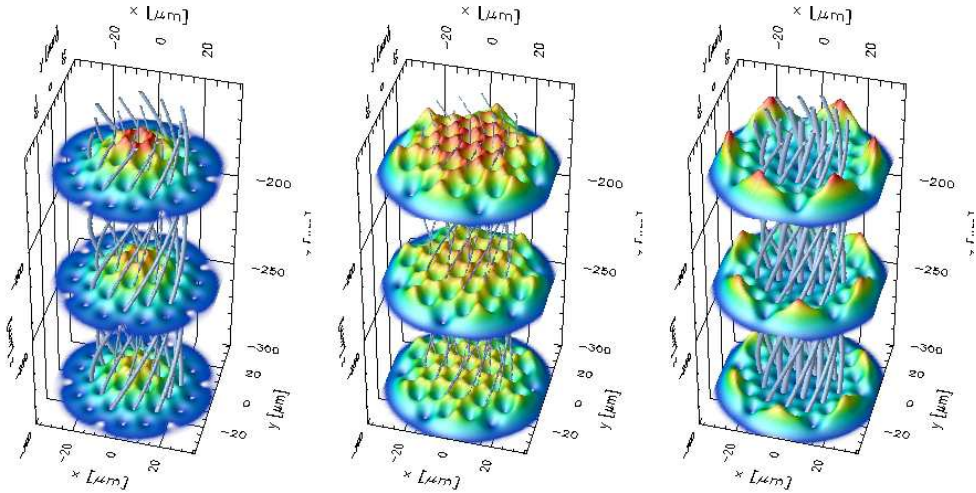}}
\caption{
Beam profile of an atom laser from a BEC with 37 vortices at different energies and heights. From the left figure to the right one we vary the detuning frequency $(-10,5,20)\,$kHz. These frequencies are also marked by dotted lines in Figure~\ref{fig:vortex37_jtot}. The density in the beam is plotted at three different distances $(177, 239, 300)\,\mu$m from the center of the BEC. The vortex cores are indicated by small tubes at all distances. BEC parameters same as Figure~\ref{fig:vortex37_jtot}.
\label{fig:vortex37_psi}}
\end{figure}

We illustrate these results using a model condensate featuring a symmetrical triangular lattice of 37 vortices $10\,\mu$m apart, embedded into a Gaussian source of width $a=5\,\mu$m.  While the frequency dependence of the outcoupling rate merely shows the familiar Gaussian character (compare Figures~\ref{fig:vortex1_jtot} and \ref{fig:vortex37_jtot}), plots of the resulting atom laser profile exhibit rich detail (Figure~\ref{fig:vortex37_psi}):  Due to the rotation of the source, the vortex pattern, which is fully transferred from the BEC into the laser beam, forms an intertwined braid-like structure along the $z$--axis.  It modulates the lateral beam profile which now strongly depends on the detuning frequency $\Delta\nu$.  The outcoupling rate varies between the different angular momentum components $\sigma_{m,\rm rf}(\mathbf r)$ (\ref{eq:Atom2.11}) that make up the source, as explained in Section~\ref{sec:Ball4} (\ref{eq:Ball4.13}).  A negative shift in the frequency suppresses states with high $|m|$, leading to an approximate Gaussian shape of the particle distribution, whereas positive detuning ($\Delta\nu > 0$) emphasizes these contributions.  The centrifugal barrier effective for them then produces a ring-like ``crown'' emission pattern.

\section{Conclusion}
\label{sec:Conc}

In our study, we established a systematic approach to scattering processes that involve non-isotropic emission of quantum particles.  Starting from the stationary Schr\"odinger equation (\ref{eq:Multi1.2}) that incorporates a source term $\sigma(\mathbf r)$ responsible for a steady particle flow, we proceeded by analogy with potential theory and introduced point-like ``multipole sources'' $\delta_{lm}(\mathbf r)$ as limiting cases of sources with $(l,m)$ orbital symmetry.  These sources, and the scattering waves and currents generated by them, are available from the conventional Dirac $\delta$--singularity and its assigned Green function $G(\mathbf r,\mathbf r';E)$ (\ref{eq:Multi1.3}) through the action of a differentiation operator of suitable spherical symmetry, the spherical tensor gradient $K_{lm}(\boldsymbol\nabla')$ (Section~\ref{sec:Multi2}).  The scattering waves emitted by multipole sources locally show pure $(l,m)$ angular symmetry, and in the absence of an external potential, they reduce to the spherical partial waves familiar from conventional scattering theory (Section~\ref{sec:Multi4}).

Notably, the local orbital characteristics are preserved also in an external potential that breaks rotational symmetry.  The multipole wave then describes the propagation of particles initially emitted in the $(l,m)$ eigenstate of angular momentum, and thus generalize the notion of a partial wave.  We performed a detailed study of the linear potential environment $U(\mathbf r) = - \mathbf r\cdot\mathbf F$, i.~e., scattering in the presence of a uniform force field $\mathbf F$.  This problem allows for an analytical solution, and closed-form expressions for the ballistic multipole waves $G_{lm}(\mathbf r,\mathbf o;E)$ (\ref{eq:Ball2.4}) and currents $J_{lm,l'm'}(E)$ (\ref{eq:Ball3.8}) are assembled in Section~\ref{sec:Ball}.  Uniformly accelerated scattering waves display a characteristic set of features, including a prominent fringe structure and a modulation of the cross section, that in a semiclassical explanation are attributed to two-path interference in the force field (Section~\ref{sec:Ball4}).

The theory of ballistic multipole waves directly applies to near-threshold photodetachment processes in an electric field environment, a topic that attracted considerable interest during the past fifteen years.  The source model immediately yields analytical expressions for the photocurrent spectrum and the spatial electron distribution that are in excellent agreement with the available experimental data.  Our attention was mainly dedicated to $p$--wave photodetachment where we discussed the influence of the laser polarization on the photoelectron current profile (Section~\ref{sec:Photo}).

Somewhat akin to the situation in potential theory, in the ballistic environment extended sources of Gaussian shape may be replaced by ``virtual'' point sources that are displaced from the center of the actual distribution.  This mapping remains feasible for ``Gaussian multipole sources'' $\sigma_{lm}(\mathbf r)$ (\ref{eq:Atom1.3}), harmonic oscillator eigenstates of $(l,m)$ orbital symmetry that are generated by the same differentiation formalism as the multipole point sources $\delta_{lm}(\mathbf r)$.  Depending on their size, these Gaussian multipoles may act as effective ballistic tunneling sources that are characterized by very different properties (Section~\ref{sec:Ball4}).  In practice, Gaussian wave functions are shared by the atoms in an ideal Bose--Einstein condensate trapped in a harmonic potential.  A ballistic tunneling source is then realized by continuous outcoupling of atoms under the influence of gravity, leading to the formation of an ``atom laser'' beam.  While a BEC in its ground state simply leads to isotropic emission, vortices embedded in the condensate will create higher angular momentum modes in the laser profile.  In particular, a BEC bearing a single vortex provides a $p$--wave ballistic tunneling source (Section~\ref{sec:Atom1}).  The particle distribution generated by it depends strongly on the relative orientation of vortex and gravitational force:  In parallel alignment, the vortex is simply preserved in the profile, while the perpendicular setup is characterized by an unusual detuning-dependent node structure in the atom distribution.  Finally, we also discussed the properties of an atom laser outcoupled from a rapidly rotating BEC sustaining a vortex lattice in Section~\ref{sec:Atom2}.  The rotating beam wave function thus created is a coherent superposition of ballistic multipole waves with very high angular momentum, and the resulting laser profile starkly depends on the detuning from resonance, while the total outcoupling rate shows little variation.

\bigskip

\noindent{\small \bf Acknowledgments.\rm \ \
We appreciate helpful discussions with C.~Blondel.  C.~B.\ gratefully acknowledges scholarships by the Alexander von Humboldt foundation and the Killam trust.  T.~K.\ received a research grant from the Leonhard--Lorenz-Stiftung.  This work was financially supported by the Deutsche Forschungsgemeinschaft project KL-315/6-1.}

\bigskip

\appendix

\section{Translation theorem for harmonic polynomials}
\label{sec:Translation}

In this appendix, we present a formula that allows to expand a harmonic polynomial $\Klm_{lm}(\mathbf r+\mathbf a)$ with shifted center into a spherical power series with respect to the origin, i.~e., in the variable $\mathbf r$.  Since the position variables $\mathbf r$ and $\mathbf a$ in the argument are interchangeable, the same series will also furnish the expansion of $\Klm_{lm}(\mathbf r+\mathbf a)$ around $\mathbf a$.  Hence, we expect the general form for this series:
\begin{equation}
\label{eq:Trans1}
\Klm_{lm}(\mathbf r+\mathbf a) = \sum_{\lambda = 0}^l \sum_{\mu = -\lambda}^\lambda C_{\lambda\mu}^{lm} \Klm_{\lambda\mu}(\mathbf r) \Klm_{l-\lambda,m-\mu}(\mathbf a) \;,
\end{equation}
where $C_{\lambda \mu}^{lm} = C_{l-\lambda,m-\mu}^{lm}$ must hold.  (Note that $\Klm_{lm}(\mathbf r)$ is a homogeneous polynomial of order $l$, so the orders in the right-hand side products of harmonic polynomials in (\ref{eq:Trans1}) must add up to $l$.  Similarly, the sum of their magnetic quantum numbers must be $m$.)  The coefficients $C_{\lambda\mu}^{lm}$ in (\ref{eq:Trans1}) have been established in closed form in Refs.~\cite{Caola1978a,Chakrabarti1995a,Bracher1999a}:
\begin{equation}
\label{eq:Trans2}
C_{\lambda\mu}^{lm} = \sqrt{\frac{4\pi(2l+1)}{(2\lambda +1)(2l-2\lambda +1)} \binom{l+m}{\lambda+\mu} \binom{l-m}{\lambda-\mu}} \;.
\end{equation}
In particular, $C_{00}^{lm} = \sqrt{4\pi}$.  (For $|m-\mu| > l-\lambda$, the coefficient vanishes.)

The general series (\ref{eq:Trans1}) simplifies if the shift in the argument of $\Klm_{lm}(\mathbf r+\mathbf a)$ takes place along the axis of quantization, i.~e., $\mathbf a = a\hat e_z$.  Then, rotational symmetry around the $z$--axis is preserved, and the quantum number $m$ is not affected by the translation.  Hence, only terms with $\mu = m$ survive in (\ref{eq:Trans1}).  Inserting the explicit value $\Klm_{\lambda0}(a\hat e_z) = \sqrt{(2\lambda+1)/4\pi}\,a^\lambda$ \cite{Messiah1964a}, we thus obtain \cite{Hobson1931a,MorseFeshbach1953a}:
\begin{equation}
\label{eq:Trans3}
\Klm_{lm}(\mathbf r+a\hat e_z) = \sum_{j = |m|}^l T_{jlm} a^{l-j} \Klm_{jm}(\mathbf r) \;,
\end{equation}
where the translation coefficient $T_{jlm}$ is given by (\ref{eq:Trans2}):
\begin{equation}
\label{eq:Trans4}
T_{jlm} = \sqrt{\frac{2l-2j+1}{4\pi}}\, C_{jm}^{lm} = \sqrt{\frac{2l+1}{2j+1} \binom{l+m}{j+m} \binom{l-m}{j-m}} \;.
\end{equation}
These developments prove useful throughout Section~\ref{sec:Ball}.

\section{Some integrals involving Airy functions}
\label{sec:appendix}

\subsection{The functions $\Qk_k(\rho,\zeta;\epsilon)$}
\label{sec:appendix1}

In the course of our investigation into ballistic multipole matter waves, integral expressions of the following type:
\begin{equation}
\label{eq:QDef}
\Qk_k(\rho,\zeta;\epsilon) = \frac {\rm i}{2\pi\sqrt\pi} \int_0^\infty
\frac{{\rm d}\tau}{({\rm i}\tau)^{k+1/2}}\, \exp\left\{{\rm i}\left[ \frac{\rho^2}\tau+\tau(\zeta-\epsilon)- \frac{\tau^3}{12} \right]\right\}
\end{equation}
are frequently encountered.  For integer indices $k$, this set of integrals permits explicit evaluation in terms of products of Airy functions.  Introducing the Airy Hankel function $\Ci(u)=\Bi(u)+ {\rm i}\Ai(u)$ \cite{Abramowitz1965a}, the basic member of this class reads:
\begin{equation}
\label{eq:Q0Explicit}
\Qk_0(\rho,\zeta;\epsilon) = \Ai(\epsilon-\zeta+\rho) \Ci(\epsilon-\zeta-\rho) .
\end{equation}
This result is easily verified by observing that (\ref{eq:QDef}) in this case reduces to the Laplace transform of the ballistic propagator in one dimension (using dimensionless units) and therefore must equal the well-known Green function of a freely falling particle in a single spatial dimension \cite{Robinett1996a}.  Indeed, apart from scaling the integrals (\ref{eq:QDef}) for positive integer indices represent the stationary ballistic Green functions in the spaces of odd dimension $D=2k+1$.  In particular, uniformly accelerated waves in physical space ($D=3$) are represented by the function $\Qk_1(\rho,\zeta;\epsilon)$ \cite{Dalidchik1976a,Bracher1998a,Kramer2002a}.

From the definition (\ref{eq:QDef}), two recurrence formulae for increasing and decreasing value of the index $k$ are immediately available:
\begin{equation}
\label{eq:QDiffForward}
\Qk_{k+1}(\rho,\zeta;\epsilon) = -\frac1{2\rho}\,\frac\partial{\partial\rho} \,\Qk_k(\rho,\zeta;\epsilon) = \left[ -\frac1{2\rho}\,\frac\partial{\partial\rho} \right]^k \Qk_0(\rho,\zeta;\epsilon),
\end{equation}
\begin{equation}
\label{eq:QDiffBack}
\Qk_{-(k+1)}(\rho,\zeta;\epsilon) = \frac\partial{\partial\zeta} \,\Qk_{-k}(\rho,\zeta;\epsilon) =
\frac{\partial^k}{\partial\zeta^k} \,\Qk_{0}(\rho,\zeta;\epsilon).
\end{equation}
(The latter equalities require $k\geq0$.)  From a practical point of view, the expressions thus obtained become rather unwieldy with growing $|k|$.  The following five-point recursion relation, again easily verified using the integral representation (\ref{eq:QDef}), presents a favorable alternative:
\begin{equation}
\label{eq:QRecursion}
\rho^2\Qk_{k+2}(\rho,\zeta;\epsilon)-
\left(k+\frac{1}{2}\right)\Qk_{k+1}(\rho,\zeta;\epsilon)
+(\zeta-\epsilon)\Qk_{k}(\rho,\zeta;\epsilon)+\frac{1}{4}\Qk_{k-2}(\rho,\zeta;\epsilon) = 0.
\end{equation}
(In passing, we remark that the general solution to (\ref{eq:QRecursion}) is given by the recurrence relations (\ref{eq:QDiffForward}) and (\ref{eq:QDiffBack}) once we replace the basic function $\Qk_0(\rho,\zeta;\epsilon)$ (\ref{eq:Q0Explicit}) by the more general expression ${\rm \tilde Q}_0(\rho,\zeta;\epsilon) = \Zi(\epsilon-\zeta+\rho)\zi(\epsilon-\zeta-\rho)$, where $\Zi(u)$, $\zi(u)$ denote two arbitrary solutions of the Airy differential equation, i.~e., linear combinations of the Airy functions $\Ai(u)$ and $\Bi(u)$.)

Finally, we inquire into the asymptotic behaviour of (\ref{eq:QDef}) in the limit $\rho \rightarrow 0$.  Here, we are interested in the case of integer index $k \geq 1$.  Then, small values of $\tau$ provide the bulk contribution to the integral, which allows us to neglect the linear and cubic terms in the exponent of (\ref{eq:QDef}).  In this approximation, the integral evaluates to a Gamma function of half-integer argument \cite{Abramowitz1965a}:
\begin{equation}
\label{eq:QAsym}
\Qk_k(\rho,\zeta;\epsilon) \sim \frac{\Gamma(k-1/2)}{2\pi^{3/2}\rho^{2k-1}} = \frac{(2k-3)!!}{2^k\pi\rho^{2k-1}} \;.
\end{equation}
Therefore, the function $\Qk_k(\rho,\zeta;\epsilon)$ diverges as $\rho \rightarrow 0$.  This singularity, however, affects only the real part of $\Qk_k(\rho,\zeta;\epsilon)$.

\subsection{The functions $\Qi_k(\epsilon)$}
\label{sec:appendix2}

Another important class of functions that regularly appears when calculating ballistic total currents is contained in (\ref{eq:QDef}) as a limiting case:
\begin{equation}
\label{eq:QiDef}
\Qi_k(\epsilon) = \lim_{\rho\rightarrow 0}\,
\lim_{\zeta\rightarrow 0}\,\Im\left\{ \Qk_k(\rho,\zeta;\epsilon) \right\}.
\end{equation}
Unlike the functions $\Qk_k(\rho,\zeta;\epsilon)$ that are divergent in this limit for $k>0$ (reflecting the multipole source singularity), their imaginary parts $\Qi_k(\epsilon)$ remain well-defined.  Obviously, $\Qi_0(\epsilon) = \Ai(\epsilon)^2$, and all other expressions are available from suitably modified recurrences (\ref{eq:QDiffForward}), (\ref{eq:QDiffBack}):
\begin{equation}
\label{eq:QiDiffForward}
{\Qi}_k(\epsilon) = \lim_{z\rightarrow0} \left\{ \left[ -\frac1{2z}\,\frac\partial{\partial z} \right]^k
\Ai(\epsilon+z)\Ai(\epsilon-z) \right\},
\end{equation}
\begin{equation}
\label{eq:QiDiffBack}
\Qi_{-k}(\epsilon) = \lim_{z\rightarrow0} \left\{
\frac{\partial^k}{\partial z^k} \Ai(\epsilon-z)^2 \right\} .
\end{equation}
$(k\geq0)$.  Thus, the functions $\Qi_k(\epsilon)$ can be extracted from the Taylor series of $\Ai(\epsilon+z)\Ai(\epsilon-z)$ and $\Ai(\epsilon-z)^2$, respectively.  For practical purposes, again the following recursion relation adapted from (\ref{eq:QRecursion}) proves more suitable:
\begin{equation}
\label{eq:QiRecursion}
\left(k+\frac{1}{2}\right)\Qi_{k+1}(\epsilon) + \epsilon\Qi_k(\epsilon) - \frac14 \Qi_{k-2}(\epsilon) = 0.
\end{equation}
For the sake of completeness, we note that unlike the functions $\Qk_k(\rho,\zeta;\epsilon)$ (\ref{eq:QDef}), the limits (\ref{eq:QiDef}) can also be evaluated for half-integer index $\Qi_{k+1/2}(\epsilon)$, which in turn allows to calculate ballistic multipole currents in spaces of even dimension.  Here, direct evaluation of the integral (\ref{eq:QDef}) shows that:
\begin{equation}
\label{eq:Qi12}
\Qi_{1/2}(\epsilon) = \frac1{2\sqrt\pi} \left\{\frac13 - \Ai_1(2^{2/3}\epsilon) \right\}
\end{equation}
holds, where $\Ai_1(u) = \int_0^u {\rm d}z\,\Ai(z)$ denotes the integral of the Airy function \cite{Abramowitz1965a}.  (In particular, $\Qi_{1/2}(0) = 1/6\sqrt\pi$.)  The other functions of half-integer index $\Qi_{k+1/2}(\epsilon)$ are available from the differentiation formula:
\begin{equation}
\label{eq:Qi12diff}
\Qi_{k-1/2}(\epsilon) = - \frac\partial{\partial\epsilon}\, \Qi_{k+1/2}(\epsilon),
\end{equation}
as well as the recursion relation (\ref{eq:QiRecursion}).

\subsection{Approximations for large values of $\epsilon$}
\label{sec:appendix3}

\begin{figure}[t]
\centerline{\includegraphics[draft=false,width=3in]{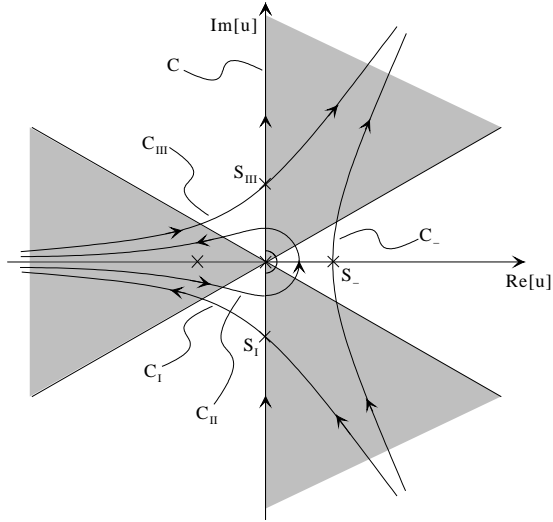}}
\caption{Evaluation of the integral (\ref{eq:Qicontour}).  The figure displays the original contour $C$ and its shifted counterparts for large values of $\epsilon$.  For $\epsilon>0$, the contour is deformed to $C_-$ and runs through the saddle point $S_-=2\sqrt\epsilon$.  In the case $\epsilon<0$, the contour is split into three parts $C_{\rm I}$, $C_{\rm II}$, $C_{\rm III}$ traversing the saddle points $S_{\rm I}$,  $S_{\rm III}$ at $\pm 2{\rm i}\sqrt\epsilon$, yet avoiding the cut in the complex $u$ plane (dark grey line) by circling the singularity at $u=0$. The integrand asymptotically vanishes in the shaded sectors. \label{fig:contour}}
\end{figure}

Next, we aim to establish the asymptotic behavior of the functions $\Qi_k(\epsilon)$ for integer $k$ in the limit $|\epsilon|\rightarrow\infty$.  The expressions thus obtained are helpful in the discussion of the ballistic multipole currents presented in section \ref{sec:Ball}.  We start out with the observation that the integral representation of these functions in (\ref{eq:QDef}) and (\ref{eq:QiDef}) may be rewritten in a complex contour integral:
\begin{equation}
\label{eq:Qicontour}
\Qi_k(\epsilon) = \frac1{4{\rm i}\pi\sqrt{\pi}} \, \int_C \frac{{\rm d}u}{u^{k+1/2}}\, {\rm e}^{-\epsilon u + u^3/12} .
\end{equation}
Here, the paths $C$ leads along the imaginary axis, avoiding the singularity at the origin and the cut in the complex plane which we choose to place onto the negative real axis (see Figure~\ref{fig:contour}).  (Note that the integrand in (\ref{eq:Qicontour}) is double-valued.  This complication is absent for half-integer indices.)

For large values of $|\epsilon|$, a saddle point approximation for (\ref{eq:Qicontour}) is in order.  This poses no problems for $\epsilon\rightarrow+\infty$ (tunneling case).  Then, the relevant stationary point of the exponent $S_-=2\sqrt\epsilon$ is readily identified, and shifting the path of integration (see Figure~\ref{fig:contour}) ultimately yields the asymptotic series:
\begin{equation}
\label{eq:Qi+large}
\Qi_k(\epsilon) \sim \frac1{2\pi} \left( 2\sqrt\epsilon \right)^{-(k+1)}
\exp\left(- \frac43 \,\epsilon^{3/2} \right) \left[ 1 - \frac{3k^2+9k+5}{24\epsilon^{3/2}} + {\cal O}\left( \frac1{\epsilon^3}\right) \right] \;.
\end{equation}
The situation is more involved for $\epsilon\rightarrow-\infty$ (classically allowed motion).  Here, the saddle points are located at $S_{\rm I,\,III} = \mp 2{\rm i}\sqrt{|\epsilon|}$, and the direction of steepest descent cuts the imaginary axis under an angle of $\pi/4$.  Hence, the integration path must be deformed to lead into the sector of asymptotically vanishing integrand $|\arg u| > 5\pi/6$.  Unfortunately, due to the presence of the cut in the complex plane, the partial paths $C_{\rm I}$, $C_{\rm III}$ cannot be simply connected as $\Re[u] \rightarrow -\infty$.  Rather, they must be linked by an additional path element $C_{\rm II}$ that loops back around the singularity located at $u = 0$, as indicated in Figure~\ref{fig:contour}.  The latter contribution is asymptotically evaluated by means of Hankel's integral formula \cite{Olver1974a} that states $\int_{C_{\rm II}} {\rm e}^t t^{-z} {\rm d}t = 2\pi{\rm i}/\Gamma(z)$:
\begin{equation}
\label{eq:QiC_II}
\int_{C_{\rm II}} \frac{{\rm d}u}{u^{k+1/2}}\, {\rm e}^{-\epsilon u + u^3/12} \sim 2\pi{\rm i} \, \frac{|\epsilon|^{k-1/2}}{\Gamma(k+1/2)} .
\end{equation}
This secular part obeys a simple power law dependence reminiscent of the Wigner law (\ref{sec:Multi4}, \cite{Wigner1948a}), while the saddle points $S_{\rm I}$, $S_{\rm III}$ deliver contributions of oscillatory character:
\begin{equation}
\label{eq:QiC_III}
\int_{C_{\rm III}} \frac{{\rm d}u}{u^{k+1/2}}\, {\rm e}^{-\epsilon u + u^3/12} \sim 2{\rm i}\sqrt\pi \left( 2{\rm i}\sqrt{|\epsilon|} \right)^{-(k+1)} \exp\left(\frac{4{\rm i}}3\, |\epsilon|^{3/2} \right) ,
\end{equation}
(cf.~eq.~(\ref{eq:Qi+large})).  The saddle point $S_{\rm I}$ adds the conjugate complex result, and upon gathering the terms in (\ref{eq:QiC_II}) and (\ref{eq:QiC_III}), one obtains the following leading asymptotic form for $\Qi_k(\epsilon)$ as $\epsilon\rightarrow-\infty$:
\begin{equation}
\label{eq:Qi-large}
\Qi_k(\epsilon) \sim \frac1{2\sqrt\pi}\, \frac{|\epsilon|^{k-1/2}}{\Gamma(k+1/2)} +
\frac1{2\pi} \left( 2\sqrt{|\epsilon|} \right)^{-(k+1)}
\sin\left(\frac43\,|\epsilon|^{3/2} - \frac{k\pi}2 \right) .
\end{equation}

\end{document}